\documentclass[apj]{emulateapj}

\def\Ha{\ifmmode^{\mathrm{H}\alpha }\else$\mathrm{H}\alpha$\fi}
\def\Hb{\ifmmode^{\mathrm{H}\beta }\else$\mathrm{H}\beta$\fi}
\def\LyA{\ifmmode^{\mathrm{H}\alpha }\else$\mathrm{Ly}\alpha$\fi}
\def\BrA{\ifmmode^{\mathrm{Br}\alpha }\else$\mathrm{Br}\alpha$\fi}
\def\BrG{\ifmmode^{\mathrm{Br}\gamma }\else$\mathrm{Br}\gamma$\fi}
\def\PaB{\ifmmode^{\mathrm{Pa}\beta }\else$\mathrm{Pa}\beta$\fi}
\def\mag{\ifmmode^{\rm m }\else$^{\rm m}$\fi}

\def\as{$\,^{\prime\prime}\,$}

\def\hh{\ifmmode^{\rm h}\else$^{\rm h}$\fi}
\def\mm{\ifmmode^{\rm m}\else$^{\rm m}$\fi}
\def\ss{\ifmmode^{\rm s}\else$^{\rm s}$\fi}
\def\deg{\ifmmode^\circ\else$^\circ $\fi}
\def\amin{\ifmmode^\prime\else$^\prime $\fi}

\def\decdm#1#2{\ifmmode{#1}\else{$#1$}\fi\deg\ #2\amin\ }

\def\dec#1#2#3{\ifmmode{#1}\else{$#1$}\fi\deg\ #2\amin\ #3\as\ }

\def\decb#1#2#3#4{\ifmmode{#1}\else{$#1$}\fi\deg\ #2\amin\ #3\farcs#4 }

\shorttitle{Resolving the gap and AU-scale asymmetries in the pre-transitional disk of V1247~Orionis}
\shortauthors{Kraus et al.}

\begin{document}


\title{Resolving  the gap and AU-scale asymmetries in the pre-transitional disk of V1247~Orionis}


\footnotetext[1]{Based on observations made 
with the Keck observatory (NASA program ID N121IV and 
NOAO program ID N121N2),
Gemini South (NOAO program ID GS-2011B-Q-19),
and with ESO telescopes at the Paranal Observatory 
(ESO program IDs 088.C-0868(A) and 088.C-0763(A+B)).        
}


\author{
Stefan Kraus\altaffilmark{1,2,3}, 
Michael J.\ Ireland\altaffilmark{4},
Michael L.\ Sitko\altaffilmark{5,6,7},
John D.\ Monnier\altaffilmark{2},
Nuria Calvet\altaffilmark{2},
Catherine Espaillat\altaffilmark{1},
Carol A.\ Grady\altaffilmark{8},
Tim J.\ Harries\altaffilmark{3},
Sebastian F.\ H\"onig\altaffilmark{9},
Ray W.\ Russell\altaffilmark{7,10},
Jeremy R.\ Swearingen\altaffilmark{5},
Chelsea Werren\altaffilmark{5}, and
David J.\ Wilner\altaffilmark{1}
}


\affil{
$^{1}$~Harvard-Smithsonian Center for Astrophysics, 60 Garden Street, MS-78, Cambridge, MA 02138, USA\\
$^{2}$~Department of Astronomy, University of Michigan, 918 Dennison Building, Ann Arbor, MI 48109, USA\\
$^{3}$~School of Physics, University of Exeter, Stocker Road, Exeter EX4 4QL, UK\\
$^{4}$~Department of Physics and Astronomy, Macquarie University, Sydney, NSW 2109, Australia\\
$^{5}$~Department of Physics, University of Cincinnati, Cincinnati, OH 45221, USA\\
$^{6}$~Space Science Institute, 475 Walnut St., Suite 205, Boulder, CO 80301, USA\\
$^{7}$~Visiting Astronomer, NASA Infrared Telescope Facility, operated by the University of Hawaii under contract with the National Aeronautics and Space Administration\\
$^{8}$~Eureka Scientific, Inc., Oakland, CA 94602; Exoplanets and Stellar Astrophysics Laboratory, Code 667, Goddard Space Flight Center, Greenbelt, MD 20771, USA\\
$^{9}$~Department of Physics, University of California Santa Barbara, Broida Hall, Santa Barbara, CA 93106, USA\\
$^{10}$~The Aerospace Corporation, Los Angeles, CA 90009, USA
}


\begin{abstract}
Pre-transitional disks are protoplanetary disks with a gapped disk structure,
potentially indicating the presence of young planets in these systems.
In order to explore the structure of these objects and their gap-opening mechanism,
we observed the pre-transitional disk \object{V1247\,Orionis} using the 
Very Large Telescope Interferometer, the Keck Interferometer, Keck-II, Gemini South, and IRTF.
This allows us spatially resolve the AU-scale disk structure
from near- to mid-infrared wavelengths (1.5 to 13\,$\mu$m), 
tracing material at different temperatures and over a wide range of stellocentric radii.
Our observations reveal a narrow, optically-thick inner-disk component (located at 0.18\,AU
from the star) that is separated from the optically thick outer disk (radii $\gtrsim 46$\,AU),
providing unambiguous evidence for the existence of a gap in this pre-transitional disk.
Surprisingly, we find that the gap region is filled with significant amounts of
optically thin material with a carbon-dominated dust mineralogy.
The presence of this optically thin gap material cannot be deduced solely from the 
spectral energy distribution, yet it is the dominant contributor at mid-infrared wavelengths.
Furthermore, using Keck/NIRC2 aperture masking observations in the $H$, $K'$, and $L'$ band, 
we detect asymmetries in the brightness distribution on scales of $\sim$ 15--40\,AU, i.e.\ within the gap region.
The detected asymmetries are highly significant, yet their amplitude and direction changes 
with wavelength, which is not consistent with a companion interpretation
but indicates an inhomogeneous distribution of the gap material.
We interpret this as strong evidence for the presence of complex density structures, 
possibly reflecting the dynamical interaction of the disk material
with sub-stellar mass bodies that are responsible for the gap clearing.
\end{abstract}





\section{Introduction}

Planets are believed to form in the circumstellar disks around young stars, 
either through a process of core accretion \citep{pol96}
or gravitational instabilities in the more extended disk regions \citep{bos00}.
A particularly interesting phase in this process starts when the newly-formed planetary bodies have
gained sufficient mass to interact with the ambient disk material and 
affect the disk structure significantly \citep[e.g.][]{par04}.
Potential candidates for disks that might have been dynamically affected by planetary bodies
are the {\it transitional disks} \citep{str89} and {\it pre-transitional disks} \citep{esp07b}.  
These objects exhibit a strong far-infrared ($\gtrsim 20$\,$\mu$m) excess, 
but have a significantly reduced near-infrared (NIR) to mid-infrared (MIR) excess compared to 
classical T~Tauri disks.
This reduced excess emission indicates that the inner-most disk regions
contain only optically thin gas and dust (transitional disks)
or exhibit an extended gap, which separates the optically thick inner disk from the outer disk
(pre-transitional disks).
The inner ``holes'' and gaps observed in transitional/pre-transitional disks 
could be caused by disk-planet interaction \citep[e.g.][]{qui04,kra12d}, 
but several alternative disk clearing scenarios have been proposed,
including grain growth \citep[e.g.][]{dul05}, 
magnetorotational instabilities \citep{chi07},
photoevaporation \citep{ale07}, and 
truncation by close-in stellar companions \citep{ire08a}.

Most of the aforementioned processes take place in the inner few astronomical units (AU)
around the central star, 
corresponding to angular scales $\rho \lesssim 0.01\,\arcsec$, even for the nearest young stars. 
Given that these scales are not accessible with conventional imaging techniques, 
earlier studies relied mostly on the modelling of spatially unresolved constraints, 
in particular the spectral energy distribution (SED). 
However, these modelling techniques suffer from well-known degeneracies -- 
grain temperature, for instance, is sensitive to both its distance from the illuminating star, 
the particle size, and the dust composition \citep{tha94,vin03}.
Therefore, we initiated an observational campaign to study these objects
using a multi-wavelength interferometry approach, 
where we combine near- and mid-infrared interferometry constraints.
The data set includes long-baseline interferometric data from the Keck Interferometer (KI)
and Very Large Telescope Interferometer (VLTI), single-dish interferometric data from Keck \& Gemini South,
and spectroscopic data from IRTF and the ESO 3.6\,m telescope.
Given the wide wavelength coverage ($1.5-13\,\mu$m), our interferometric observations probe 
a wide range of dust temperatures and material located over a wide range of stellocentric radii,
from sub-AU to tens of AU.

Here, we report first results from our campaign, focusing on the 
pre-transitional disk object \object{V1247\,Orionis} (HD\,290764, PDS\,192).
V1247\,Ori is likely a member of the Orion~OB1\,b association
\citep{sch71,gue81} and the Alnilam cluster (also known as 
``$\epsilon$~Orionis cluster'' and ``Collinder 70''),
whose age has been estimated to 5-10\,Myr \citep{cab08b}.
There is also considerable debate in the literature concerning the precise 
spectral classification of V1247\,Ori, ranging from spectral types 
of F0V \citep{vie03} to A5III \citep{sch71}, which has caused us to
re-evaluate the spectral classification in our study.
\citet{cab10} reported the observation of two isolated, but deep UX\,Ori-like occultation events, 
which were attributed to occultation from disk material in the inner disk regions.
The SED of V1247\,Ori shows a significantly reduced excess emission
in the wavelength range $3-15$\,$\mu$m \citep{cab10},
revealing similarities with other pre-transitional disk objects.
However, V1247\,Ori is of considerably earlier spectral type than
most other pre-transitional disks, which makes this system an
interesting laborary to study disk evolution and planet-formation
in the intermediate-mass regime ($\sim 2\,M_{\sun}$). 

In the following, we present our extensive observational data set (Sect.~\ref{sec:observations})
and our results, including a re-evaluation of the spectral classification of the central star (Sect.~\ref{sec:resultsspecclass}),
the SED and line diagnostics (Sect.~\ref{sec:resultsdiskSED}),
our interferometric constraints on the disk structure (Sect.~\ref{sec:resultsdisk}),
and a global model based on our multi-wavelength interferometry and SED constraints (Sect.~\ref{sec:modeling}).
Finally, we discuss the implications of our work on the understanding of the pre-transitional disk phase 
(Sect.~\ref{sec:discussion}) and summarize our conclusions (Sect.~\ref{sec:conclusions}).

\section{Observations}
\label{sec:observations}

\begin{deluxetable*}{lccccc}
\tablecolumns{4}
\tablewidth{0pc}
\tablecaption{Observation log\label{tab:obslog}}
\tablehead{
\colhead{Instrument} & \colhead{UT Date} & \colhead{Mode} & \colhead{Seeing} & \colhead{Calibrator}\\
& [YYYY/MM/DD]  &   &  [arcsec] & }

\startdata
IRTF/SpeX &  2010/02/28 & SXD &  0.6 & HD\,34203\\
                      && LXD & & HD\,34203\\ 
                      && Prism & & HD\,34203\\
IRTF/SpeX\tablenotemark{a} &  2011/03/04  & SXD & 0.8 & HD\,34203 \\
IRTF/BASS\tablenotemark{b} & 2011/10/17   &     & 0.7 & HD\,29139 \\
IRTF/SpeX\tablenotemark{c} &  2011/10/18  & SXD & 0.6 & HD\,34203 \\
                     && LXD &  & HD\,34203 \\ 
 KI/ASTRA &  2011/11/05  & V2-SPR & 0.6 & HD\,41794, HD\,37331 \\
 Gemini\,S/T-ReCS &  2011/11/23  & Si-2 & 0.5 & HD\,36167 \\
                     && Si-3  & & HD\,36167 \\ 
                     && Si-5  & & HD\,36167 \\ 
 VLTI/AMBER &  2011/12/09  & LR-$HK$  &  1.1 & HD\,37331, HD\,36059 \\
 VLTI/MIDI &  2011/12/11  & Prism  &  0.9 & HD\,39400 \\
 VLTI/MIDI &  2011/12/12  & Prism &  0.7 & HD\,39400 \\
 Keck-II/NIRC2 &  2012/01/08  & K' & 0.4 & HD\,37634, HD\,38406 \\
 Keck-II/NIRC2 &  2012/01/10  & H  & 1.1 & HD\,37331 \\
                             && L' & 1.3 & HD\,37634, HD\,38406 \\ 
 IRTF/SpeX &  2012/02/26 & SXD   & 0.9 & HD\,34203\\ 
                        && LXD   &     & HD\,34203\\ 
                        && Prism &     & HD\,34203\\
IRTF/SpeX &  2012/09/12  & SXD   & 0.8 & HD\,34203\\
                        && LXD   &     & HD\,34203\\ 
                        && Prism &     & HD\,34203\\
IRTF/SpeX &  2012/11/04  & SXD   & 0.4 & HD\,34203\\
                        && LXD   &     & HD\,34203\\
                        && Prism &     & HD\,34203
\enddata
\tablenotetext{a}{SXD only due to clouds (1-mag variations in sky transparency).}
\tablenotetext{b}{Exceptionally low precipitable water vapor allowed observations through the 6 $\mu$m water vapor band.}
\tablenotetext{c}{No Prism observations due to thin cirrus.}
\end{deluxetable*}

\subsection{IRTF/SpeX+BASS near- and mid-infrared spectroscopy and archival spectro-photometry}

\begin{figure*}[t]
  \centering
  \includegraphics[angle=0,scale=0.65,trim=0mm 0mm 0mm 8mm,clip]{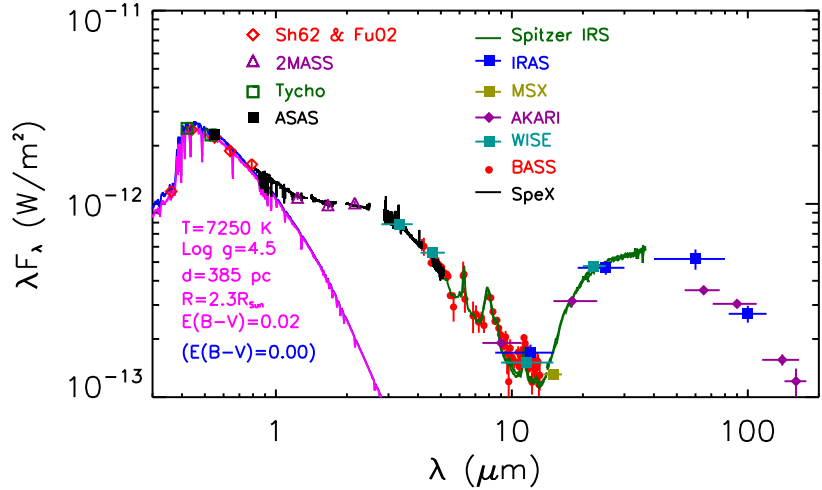}  
  \caption{
    SED of V1247\,Ori, compiled from our IRTF/SpeX (epoch 2010 February 28) and 
    IRTF/BASS (epoch 2011 January 17) observations and data from the literature.
    The horizontal error bars indicate the bandpass of 
    the employed photometric filters.  The photospheric emission (blue and magenta curve) 
      is represented with the SYNTHE model \citep{kur93} for a F0V star 
      with $T_{\rm eff}=7250$\,K and $\log g=4.5$.
  }
  \label{fig:SED}
\end{figure*}

An important first step is to build the SED of V1247\,Ori in the relevant 
near- to mid-infrared regime.  For this purpose, we employed 
the SpeX spectrograph \citep[1--5\,$\mu$m;][]{ray03} and The Aerospace Corporation's 
Broad-band Array Spectrograph System (BASS; 3--14\,$\mu$m), which are mounted at 
NASA's Infrared Telescope Facility (IRTF).  The observing dates and observing modes
are summarized in Table~\ref{tab:obslog}, together with our other observations.

The SpeX spectra were recorded using the echelle grating in both short-wavelength mode 
(SXD, 0.8--2.4\,$\mu$m) and long-wavelength mode (LXD, 2.3--5.4\,$\mu$m) using a {0.8\arcsec} slit.
The spectra were corrected for telluric extinction and flux calibrated 
against the calibrator star \object{HD\,34203}
using the Spextool data reduction package \citep{vac03,cus04}.
In addition to the {0.8\arcsec}-slit spectra, we also recorded data with
a wide {3\arcsec} slit, which allows us to retrieve the absolute flux levels.
V1247\,Ori was observed six times in total with SpeX (see Table~\ref{tab:obslog}), 
allowing us to detect possible mid-infrared variability.

Our BASS short-wavelength (2.9--6\,$\mu$m) and long-wavelength 
(6--13.5\,$\mu$m) observations cover the full $L$, $M$, and $N$ bands
with a spectral resolution ranging from about 30 to 125.
For extracting the BASS data, we used the Spextool data reduction package.
The calibrator star was \object{HD\,29139}.

The $UBVRI$ photometry was taken from the compilation by \citet{cab10},
which contains measurements from \citet{sha62}, \citet{fuj02},
the {\em Infrared Astronomical Satellite} \citep[{\em IRAS},][]{ira88}, 
the {\em Midcourse Space Experiment} \citep[{\em MSX},][]{ega03}, 
and the {\em Akari} observatory \citep{ish10}.
In addition, we added $JHK$ band photometry from {\em 2MASS} \citep{skr06}.
Besides the ASAS $V$-band data published in \citet{cab10}, no new
ASAS observations are available.
The {\it Spitzer} IRS data were obtained from the Spitzer heritage Archive 
Post-Basic Calibrated Data (PBCD). All spectra were low resolution spectra, 
with the deviant end points of each spectral segment trimmed off. 
The resulting SED is shown in Fig.~\ref{fig:SED}.

\subsection{KI near-infrared interferometry}

The Keck Interferometer allowed us to 
resolve the distribution of the hot ($T \gtrsim 1000$~K)
circumstellar material around V1247\,Ori, forming
an effective projected baseline length $B_{p}$ of 70.1\,m
towards position angle (PA, measured East of North) {27\deg}.
The observation was conducted on 2011 November 5 
and used the ASTRA beam combination instrument \citep{pot09} 
in its self-phase-referencing mode (V2-SPR),
covering the full $K$ band with spectral resolution
$R=\lambda/\Delta\lambda=2000$.
The science target observations were accompanied by
observations of the calibrator stars
\object{HD\,41794} and \object{HD\,37331},
which allowed us to correct for atmospheric and instrumental effects.
For the calibrator sizes we adopt uniform disk (UD) diameters
of $0.213 \pm 0.015$~milliarcsecond (mas) for HD\,41794 (computed with ASPRO)
and $0.144 \pm 0.1$~mas for HD\,37331 (computed with ASPRO).

From these observations, we derived a spectrum as well as
wavelength-differential visibilities and phases.
However, neither the spectrum nor the interferometric 
observables show any significant spectral features
or wavelength-differential features.
Therefore, we use only the recorded broadband visibility information
for the modeling described in the following sections.

\subsection{VLTI/AMBER near-infrared interferometry}

\begin{figure}[t]
  \centering
  \includegraphics[angle=0,scale=0.75]{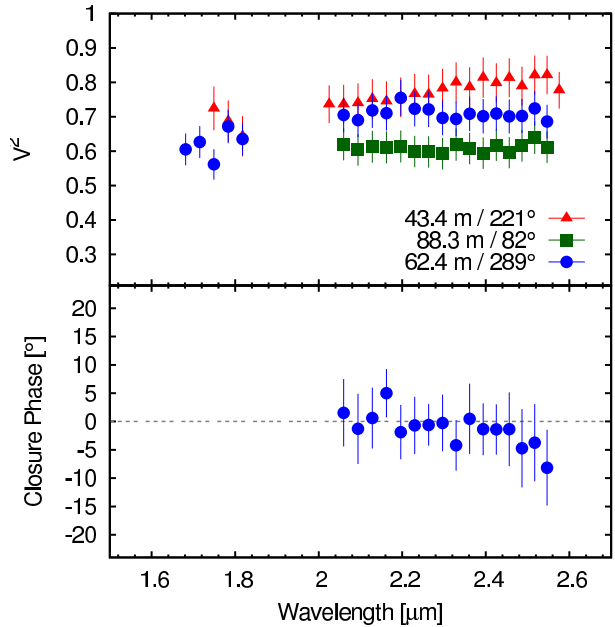}  
  \caption{
    Squared visibility amplitudes (top) and closure phases (bottom)
    in the $H$- and $K$ band measured with VLTI/AMBER.
  }
  \label{fig:VISAMBER}
\end{figure}

Additional NIR long-baseline interferometric observations
were obtained on 2011 December 9 with the AMBER instrument \citep{pet07},
improving the $uv$-plane coverage of the KI interferometry.
Using the VLTI 8.2\,m unit telescope (UT) triplet UT2-UT3-UT4,
the observations sample projected baseline lengths (position angles) of
43.4\,m (221\deg), 62.4\,m (289\deg), and
88.3\,m (82\deg).  Using optical fibers, AMBER provides 
spatial filtering, resulting in an effective field-of-view of 60\,mas.
For the beam combination we used AMBER's LR-$HK$-mode,
which provides a spectral dispersion of $R=35$
and covers wavelengths between 1.5 and 2.5\,$\mu$m 
($H$- and $K$ band).
We recorded 5000 interferograms with an integration time of 26\,ms
for V1247\,Ori and the calibrators \object{HD\,37331}
and \object{HD\,36059} (UD $0.549 \pm 0.038$~mas, ASPRO).
Data from another calibrator (\object{HD\,36811}) were rejected, 
since this object was found to be a close binary 
system\footnote{Given that our companion detection for \object{HD\,36811} might be
of interest for science programs on Am-type stars, 
we determined the binary parameters and obtained a separation 
$\rho=2.0 \pm 0.2$\,mas with position angle PA=$80 \pm 2^{\circ}$ (epoch 2011-12-09) 
and flux ratio $F_{B}/F_{A} = 0.44 \pm 0.01$ in the $H$/$K$ band.}.

From the data we extract wavelength-dependent visibilities 
and closure phases using the {\em amdlib} software \citep[Release 3;][]{tat07b,che09}.
In order to minimize the effect of residual telescope
jitter, we follow the standard AMBER data reduction
procedure and select the interferograms with the 
10\% best signal-to-noise ratio.
In addition, we reject scans where the optical path delay
exceeds 4\,$\mu$m.
The wavelength calibration was done using the telluric
absorption bands between the $J$/$H$ and $H$/$K$ band
\citep[see Appendix~A in][]{kra09a}.
The final calibrated observables are shown in Fig.~\ref{fig:VISAMBER}.

\subsection{VLTI/MIDI mid-infrared interferometry}

\begin{figure}[t]
  \centering
  \includegraphics[angle=270,scale=0.35]{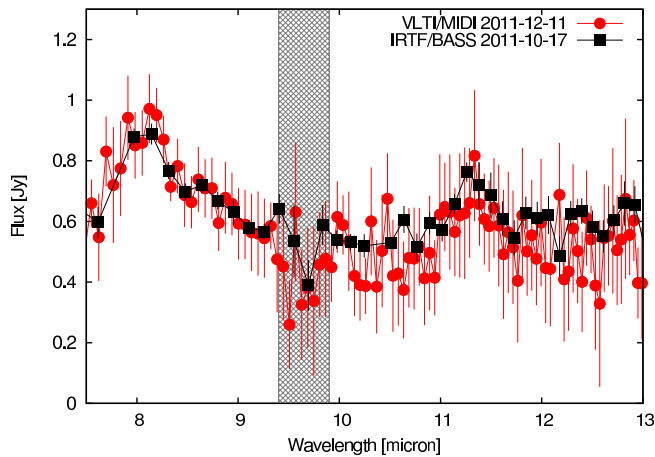}   
  \caption{
    Comparison of our VLTI/MIDI spectrum with the mid-infrared
    spectrum recorded with the IRTF/BASS.  The shaded area
    marks the location of the atmospheric ozone absorption bands,
    which might introduce additional scatter in the spectrum.
  }
  \label{fig:MIDIspectrum}
\end{figure}

\begin{figure}[t]
  \centering
  \includegraphics[angle=270,scale=0.35]{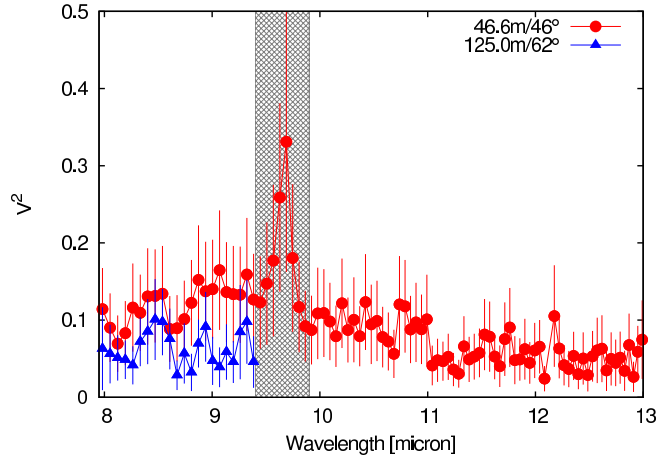}   
  \caption{
    Visibilities derived from our VLTI/MIDI observations.
    The shaded area marks the location of the atmospheric ozone absorption bands,
    which might introduce additional scatter in the spectrum.
  }
  \label{fig:MIDIVIS}
\end{figure}

Our mid-infrared long-baseline interferometric observations
were obtained with the VLTI/MIDI instrument \citep{prz03}.
The employed telescope baselines were
UT2-UT3 (2011 December 11, $B_{p}=46.6$\,m, PA=$46^{\circ}$) and 
UT1-UT4 (2011 December 12, $B_{p}=125.0$\,m, PA=$63^{\circ}$), 
i.e.\ probing similar position angles, but significantly
different spatial frequencies.
The observations on V1247\,Ori were bracketed with
observations of the calibrator star
HD\,39400 (UD $2.387 \pm 0.138$~mas, computed using CalVin).

In order to extract the mid-infrared correlated fluxes, we employ the 
MIA+EWS-assisted \citep{jaf04,lei04} faint-source 
reduction procedures described in \citet{kis11}.
Together with the MIDI interferograms, we also recorded 
photometry files on the science star and calibrator,
which allowed us to retrieve the absolute flux photometry.
Comparing the MIDI total flux spectrum with a BASS
spectrum recorded two months earlier (Fig.~\ref{fig:MIDIspectrum})
reveals that the spectra are fully consistent within the
measurement uncertainties and we find no indications
for mid-infrared flux variability.
Therefore, we decided to use the BASS spectrum in order
to convert the measured correlated fluxes to visibility amplitudes
(Fig.~\ref{fig:MIDIVIS}).

The wavelength range covered by MIDI contains also two
hydrocarbon features (Sect.~\ref{sec:resultsdiskSED}). 
However, we detect no significant change in the visibility at the 
wavelength of these organic bands (Fig.~\ref{fig:MIDIVIS}),
suggesting that the hydrocarbon grains are located on similar 
spatial scales as the optically thin dust grains.
However, observations with shorter baselines and with higher signal-to-noise ratio
will be required for a detailed quantiative analysis.

\subsection{Gemini/T-ReCS mid-infrared speckle interferometry}
\label{obs:TReCS}

\begin{figure}
  \centering
  $\begin{array}{c@{\hspace{5mm}}c}
    \includegraphics[angle=90,scale=0.5]{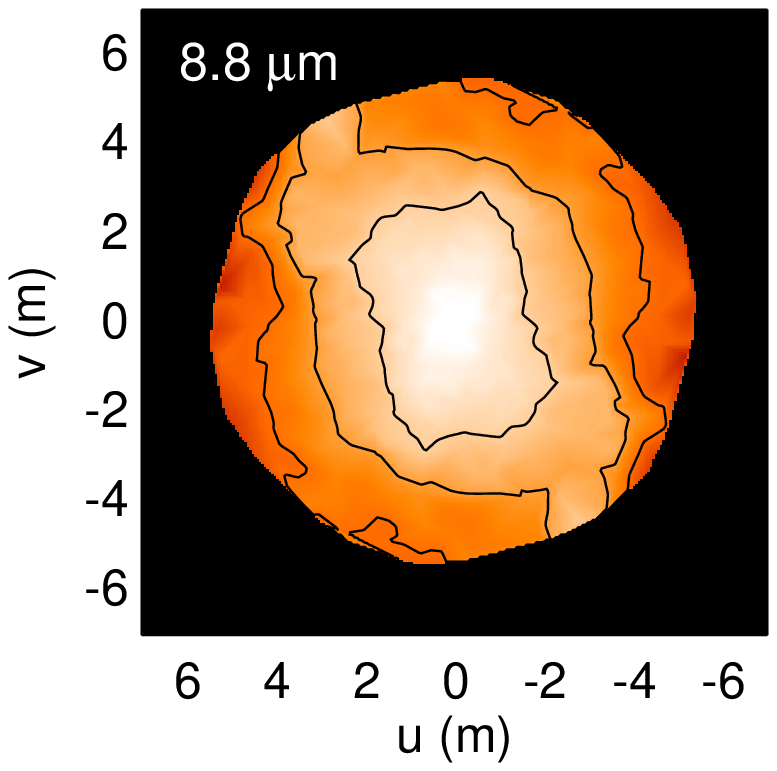} &   
    \includegraphics[angle=90,scale=0.5]{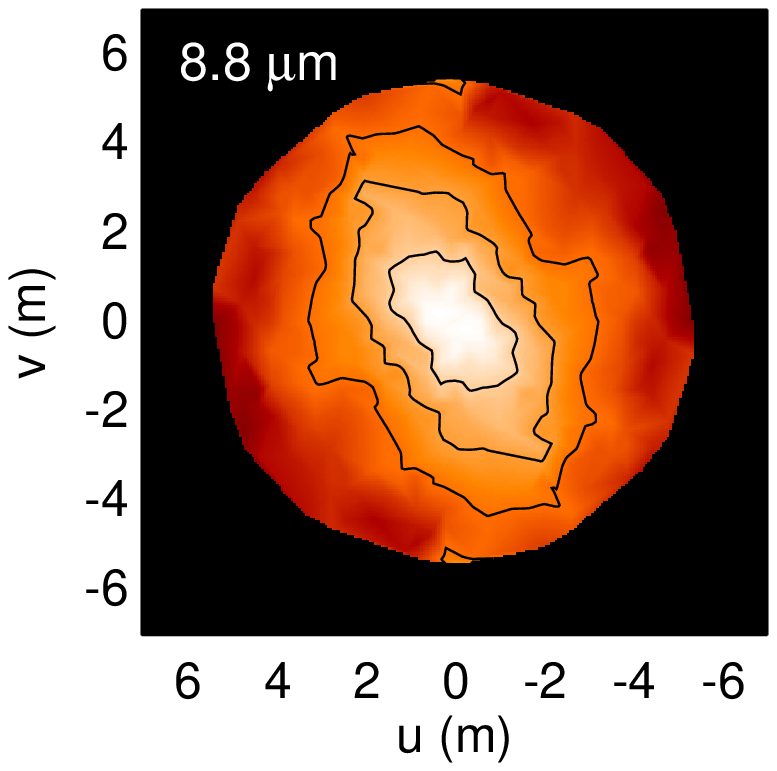} \\  
    \includegraphics[angle=90,scale=0.5]{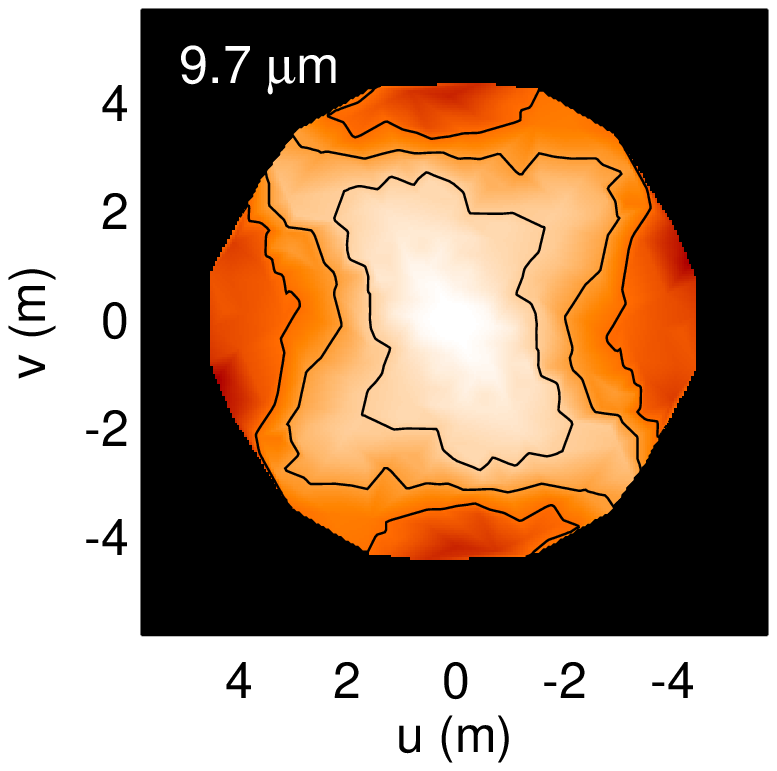} &   
    \includegraphics[angle=90,scale=0.5]{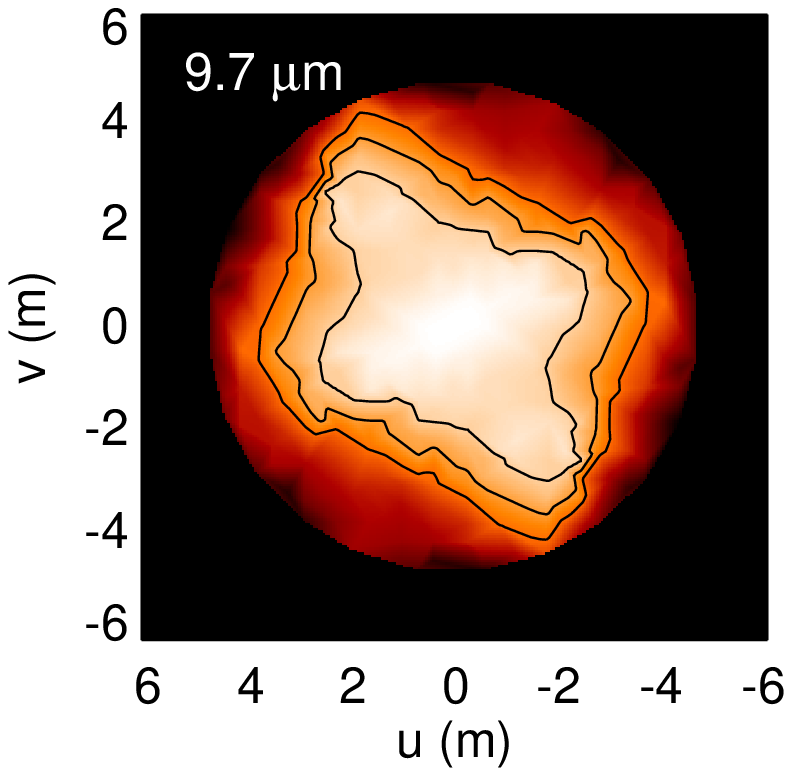} \\  
    \includegraphics[angle=90,scale=0.5]{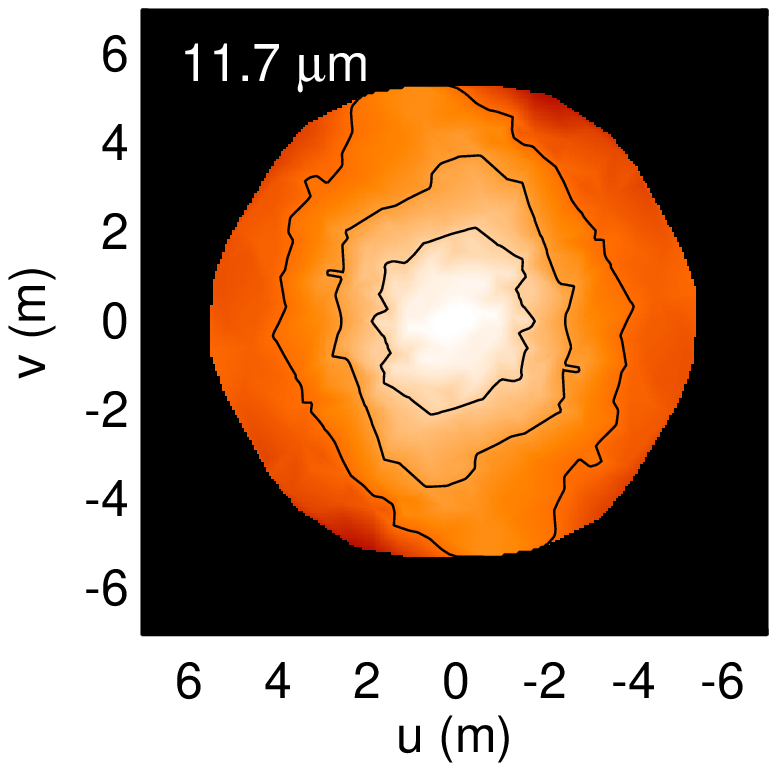} &   
    \includegraphics[angle=90,scale=0.5]{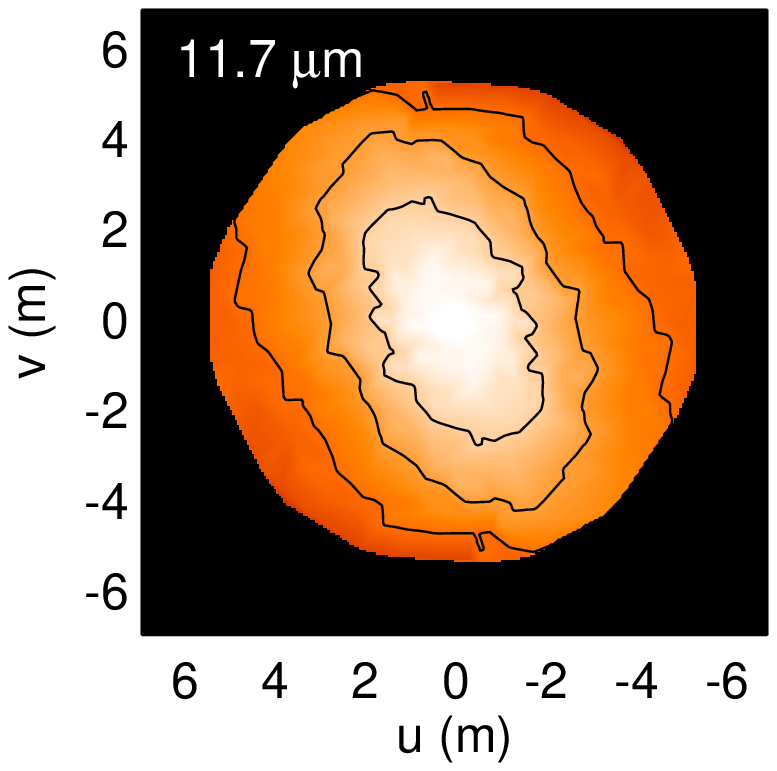}     
  \end{array}$
  \caption{
    2-D power spectra, derived from our Gemini/T-ReCS speckle observations using wavelength bins
    around 8.74\,$\mu$m (Si-2, top), 9.69\,$\mu$m (Si-3, middle), 
    and 12.66\,$\mu$m (Si-5, bottom).
    For each filter we recorded two calibrated data sets on V1247\,Ori (left and right),
    which confirm the detected object elongation independently.
    The contours mark visibility levels of 0.7, 0.8, and 0.9 in the derived power spectra.
    North is up and East is left.
  }
  \label{fig:TReCSpowerspec}
\end{figure}

Our MIDI observations show that V1247\,Ori is already strongly resolved 
on the shortest VLTI UT baseline ($B_p=46.6$\,m), indicating the presence
of an extended emission component that is not properly constrained by
our long-baseline interferometric observations.
Therefore, we conducted complementary observations using the
T-ReCS mid-infrared imager \citep{tel98,deb05} that is mounted on the Gemini/South 8.2\,m telescope.
Employing short integration times of 0.2\,s and 
a speckle interferometry analysis approach
allow us to effectively freeze the atmospheric perturbations and to 
extract interferometric visibilities and phases for effective baseline lengths $B_p \lesssim 5.5$\,m
that can be directly combined with our long-baseline interferometric data.
The data were recorded on UT 2012 November 23
under exceptional atmospheric
conditions using the T-ReCS narrowband filters
Si-2 ($\lambda_c = 8.74\,\mu$m, $\Delta\lambda=0.39\,\mu$m, where
$\lambda_c$ denotes the central wavelength of the filter and $\Delta\lambda$ the spectral bandwidth),
Si-3 ($\lambda_c = 9.69\,\mu$m, $\Delta\lambda=0.46\,\mu$m), and
Si-5 ($\lambda_c = 11.66\,\mu$m, $\Delta\lambda=0.56\,\mu$m).
Compared to the other two channels, the Si-3 data
exhibit a reduced signal-to-noise ratio, reflecting the 
low flux of the object at this wavelength
and the presence of the atmospheric ozone absorption bands.
As a calibrator star, we employed \object{HD\,36167}.

The power spectrum analysis method is described in more
detail in \citet{eis09}. 
Speckle observations in the mid-infrared are rather
sensitive to variations in the thermal background,
which can potentially induce a bias in the 
absolute visibility level.
In order to correct this problem, we renormalize the
measured visibility profiles on the shortest spatial frequencies.

The derived 2-D visibilities show consistently an
object elongation (Fig.~\ref{fig:TReCSpowerspec})
that we will further quantify in Sect.~\ref{sec:resultsdiskMIR}
using model-fitting techniques.
The validity of this elongation is also confirmed
by the fact that the T-ReCS data were recorded with
deactivated field rotator, which would introduce
a significant PA rotation in the derived 
2-D visibilities for any telescope-induced artefacts.

\subsection{Keck-II/NIRC2 near-infrared aperture masking}

\begin{figure}[tbp]
  \centering
  \includegraphics[angle=0,scale=0.55]{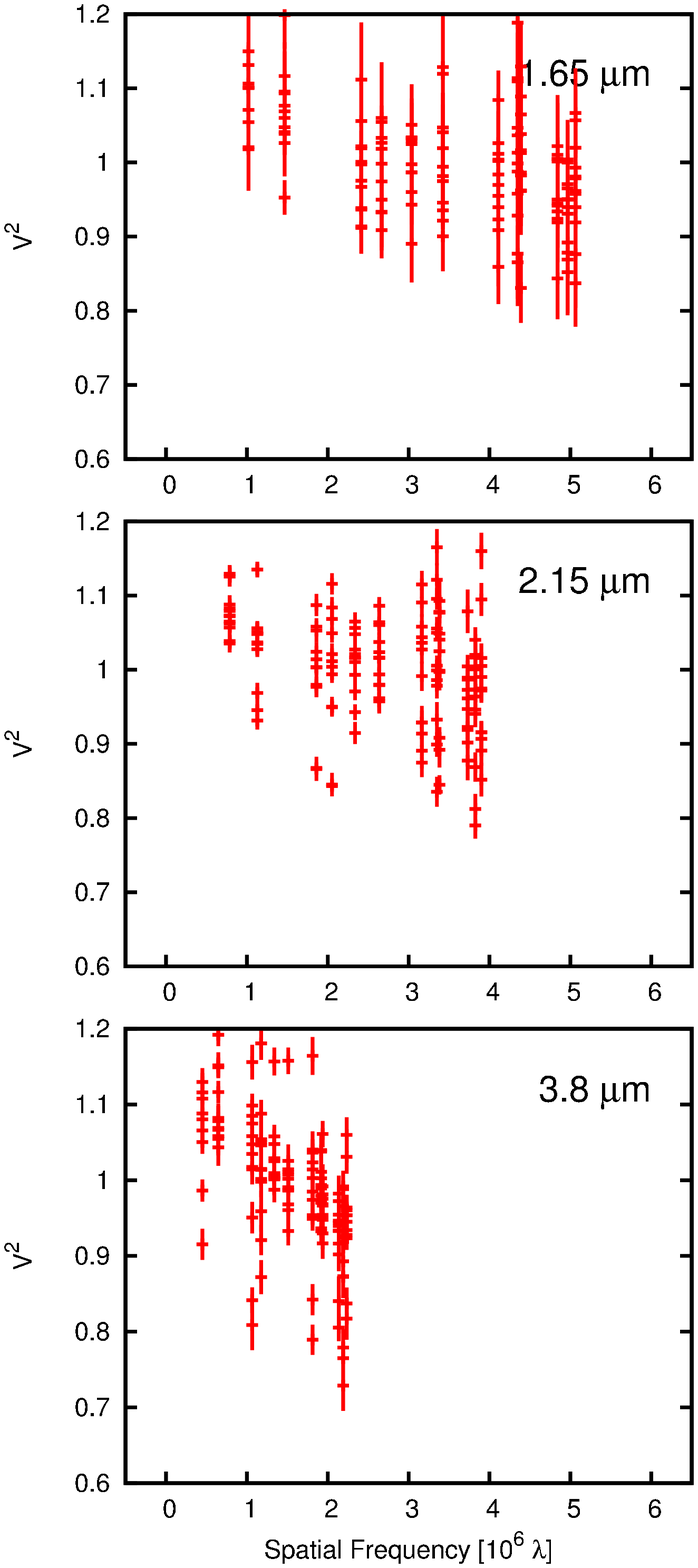}
  \caption{
    Squared visibility amplitudes derived from our Keck/NIRC2 aperture masking observations.
  }
  \label{fig:visprofile-NIRC2}
\end{figure}

\begin{figure*}[t]
  \centering
  $\begin{array}{c@{\hspace{-2mm}}c@{\hspace{-2mm}}c}
    \includegraphics[angle=0,scale=0.45]{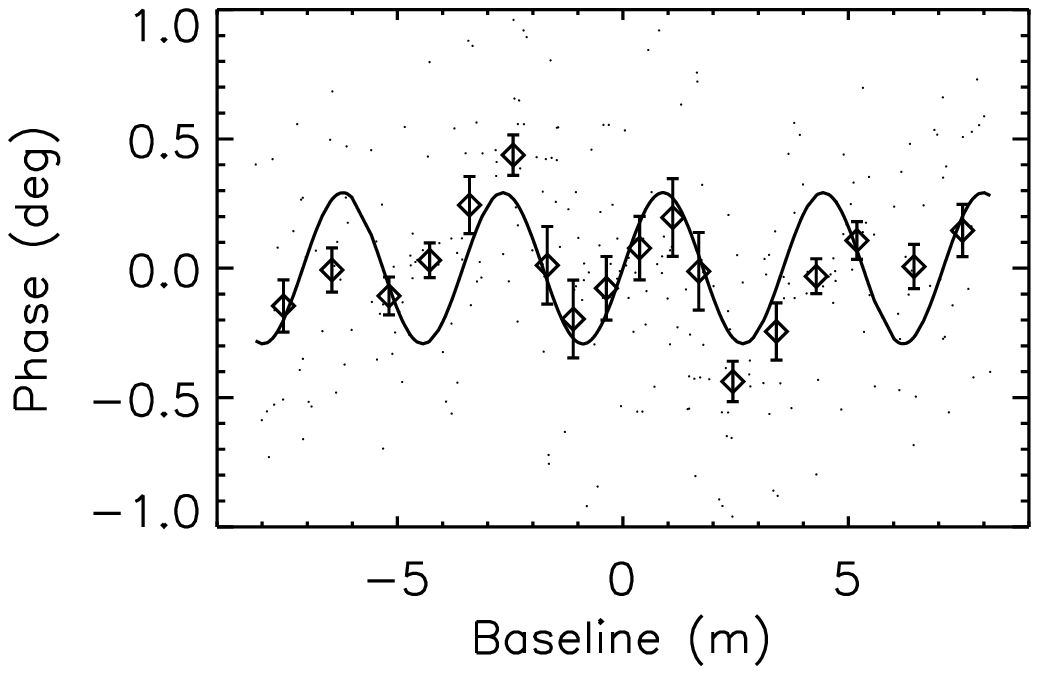} &
    \includegraphics[angle=0,scale=0.45]{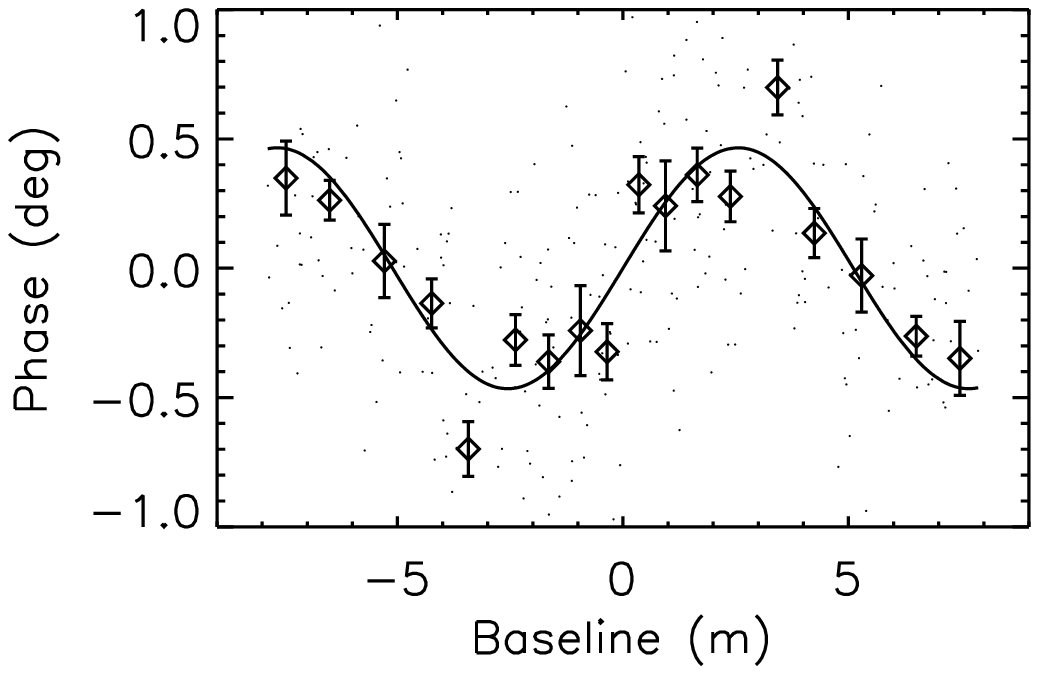} &
    \includegraphics[angle=0,scale=0.45]{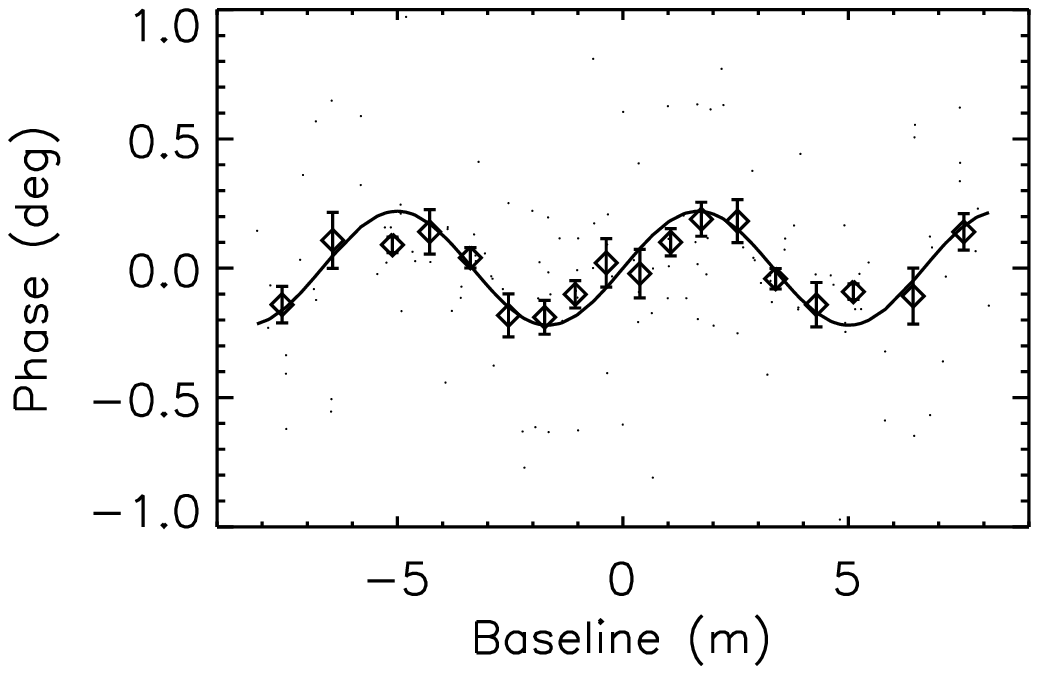} \\
    \includegraphics[angle=0,scale=0.47]{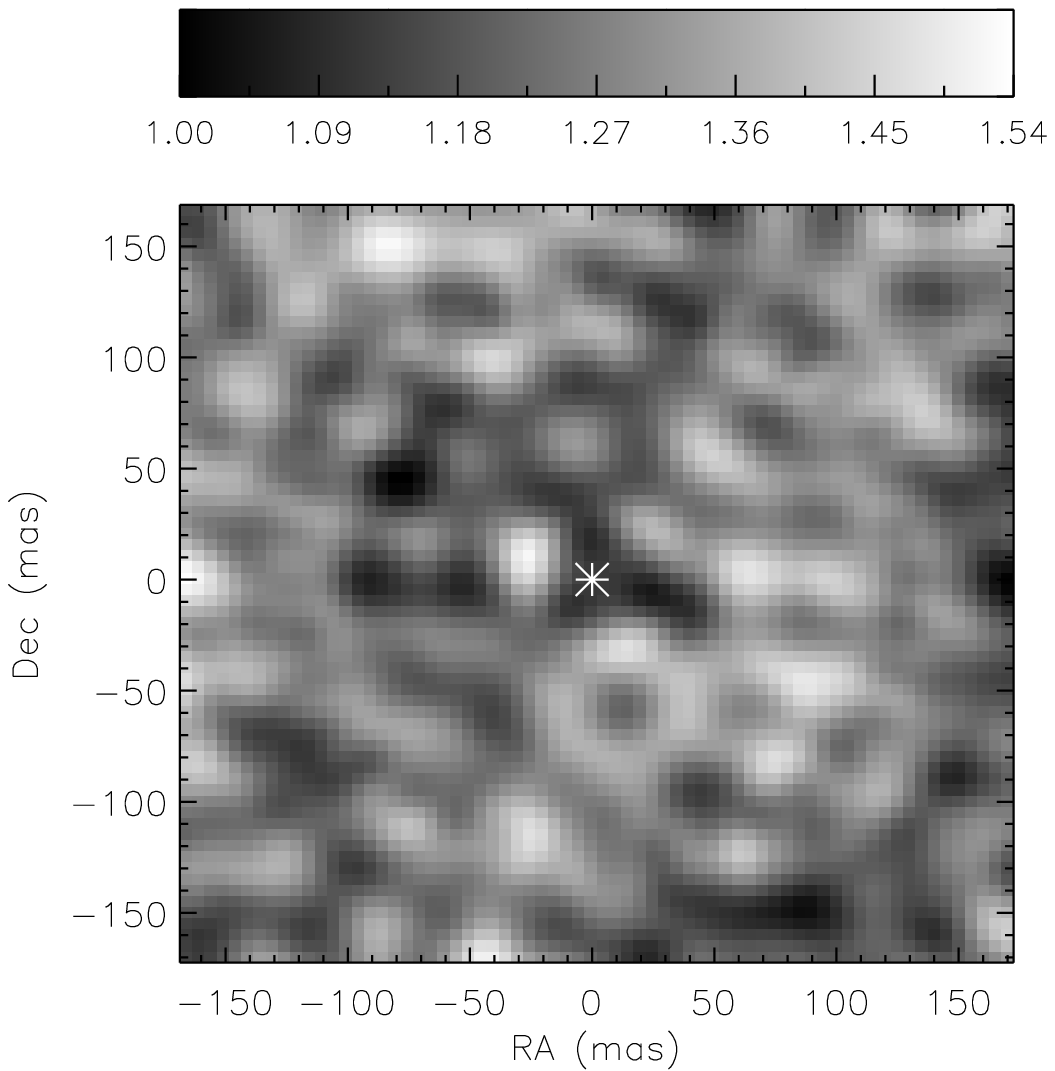} &
    \includegraphics[angle=0,scale=0.47]{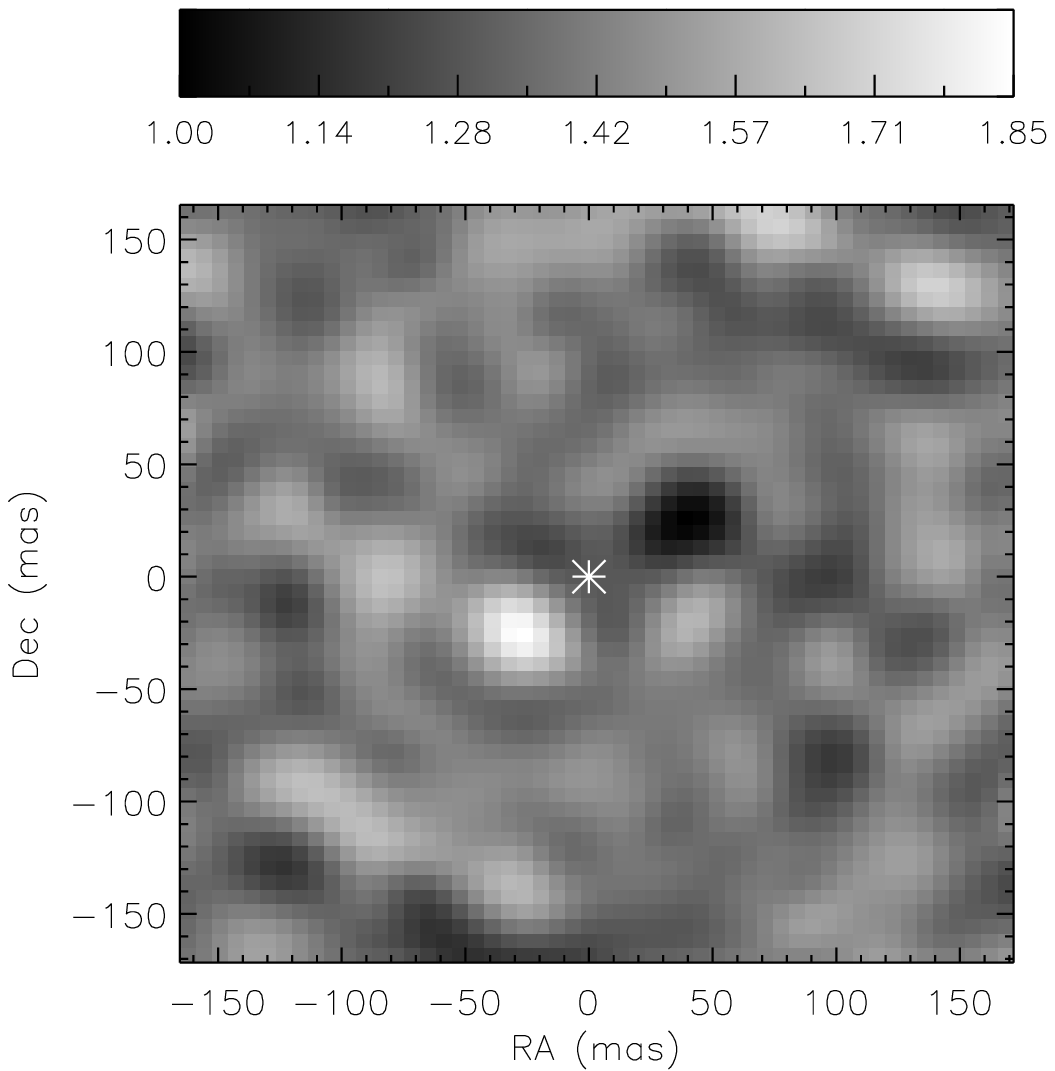} &
    \includegraphics[angle=0,scale=0.47]{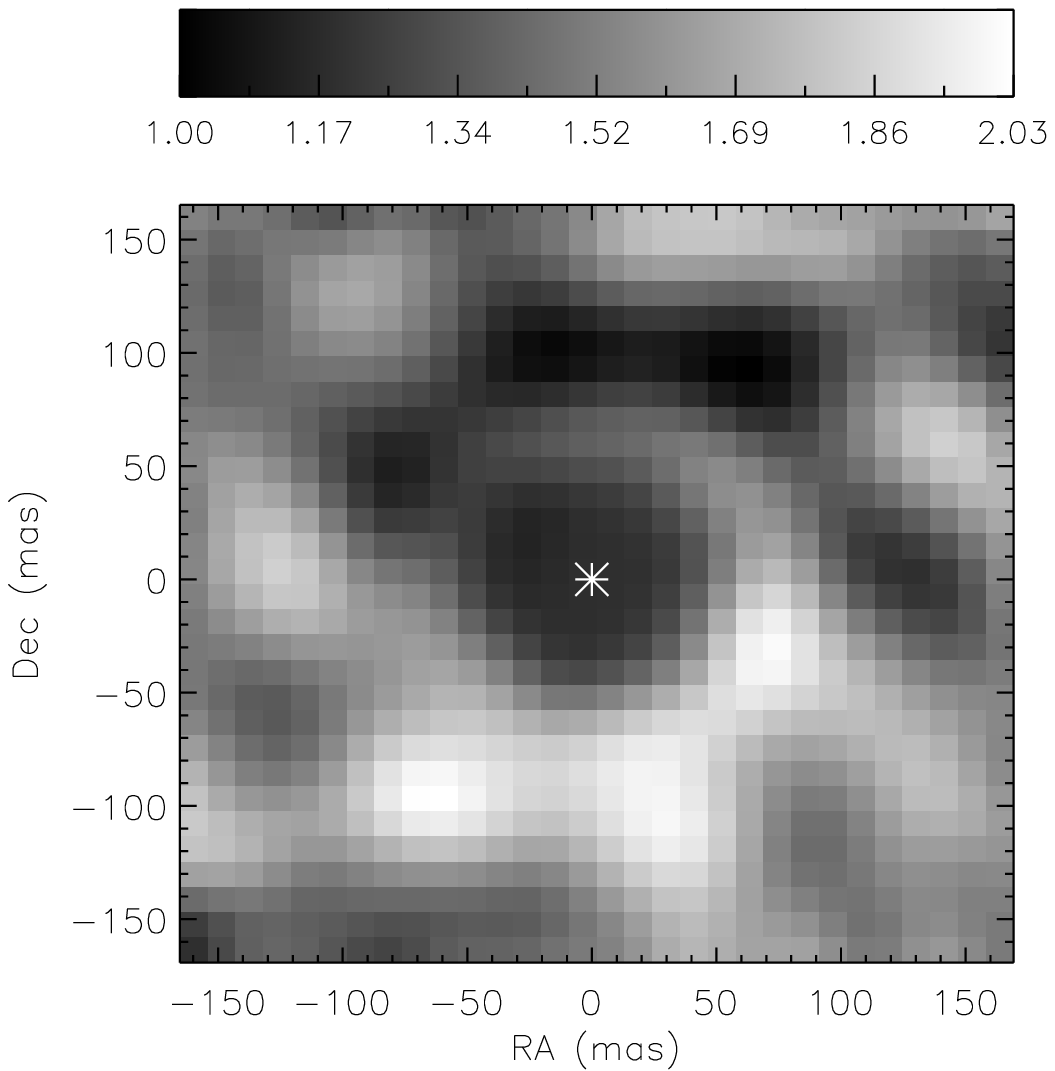}  
  \end{array}$
  \caption{
    {\it Top:} Our Keck/NIRC2 $H$-band (left column), $K'$-band (middle column), and $L'$-band (right column) aperture masking observations 
    reveal strong non-zero phases that we fit with a companion model (solid line).
    The 2-D Fourier phases have been projected perpendicular to the position angle of the best-fit companion model.
    {\it Bottom:} $\chi^2$-maps derived by fitting a companion model.  The best-fit companion positions do not agree for the
    different wavelength bands, indicating that the measured asymmetries are not caused by a simple point source, but
    complex disk asymmetries (Sect.~\ref{sec:resultsNIRC2}).
  }
  \label{fig:phases-NIRC2}
\end{figure*}

Using spatial filtering by single-mode fibers, our AMBER observations are insensitive 
to structures located at separations $\rho$ larger than $\sim 30$\,mas from the central star.
In order to probe larger spatial scales, we employed
aperture masking observations with the Keck-II 10\,m telescope
and the NIRC2 instrument.
Using a non-redundant 9-hole pupil mask allows us to derive high-precision 
interferometric phases that are well-suited to search for faint companions.
The observations were conducted on 2012 January 8 and 10 
with an integration time of 1\,s and with the adaptive optics loop closed.
The observations on 2012 January 8 were conducted 
under excellent atmospheric conditions using a $K'$-band (1.96--2.29\,$\mu$m) filter.
We executed an observing sequence of four target-calibrator pairs
with a total on-source integration time of 960\,s.
On January 10, four additional sequences were recorded using an 
$H$-band (1.5-1.8\,$\mu$m), $L'$-band (3.43-4.13\,$\mu$m), 
and again a $K'$-band filter.  
The observed calibrator stars were
\object{HD\,37331} ($H$ band),
\object{HD\,37634} ($K'$/$L'$ band), and 
\object{HD\,38406} ($K'$/$L'$-band).

The NIRC2 data were reduced using a data reduction pipeline
that was used already in various earlier studies \citep[e.g.][]{ire08a,kra12d},
providing absolute calibrated visibilities (Fig.~\ref{fig:visprofile-NIRC2})
and closure phases (Fig.~\ref{fig:phases-NIRC2}, top).

\subsection{Archival HARPS visual spectroscopy}

Given the considerable uncertainties on the 
spectral type of V1247\,Ori, we obtained HARPS high-resolution 
spectra from the ESO archive.
The 37 exposures were reduced with the HARPS DRS pipeline v3.0 and provide 
a resolution $R=120,000$.
The data were recorded between 2008-11-11 and 2008-11-13 in the course 
of a radial velocity survey (ESO program ID 082.D-0833(A), PI: Boehm).
Based on this original science objective, the data were recorded without 
associated telluric calibrators.  Therefore, we were not able to remove atmospheric 
features and thus focused our analysis on the spectral regions around 380-460\,nm 
and around the H$\alpha$ line, which suffers only from very weak telluric contamination.
The individual spectra were averaged and the continuum-level was then normalized
by fitting a higher-order polynomial.

\section{Results}
\label{sec:results}

\subsection{Spectral classification}
\label{sec:resultsspecclass}

\begin{figure}[t]
  \centering
  \includegraphics[angle=270,scale=0.35]{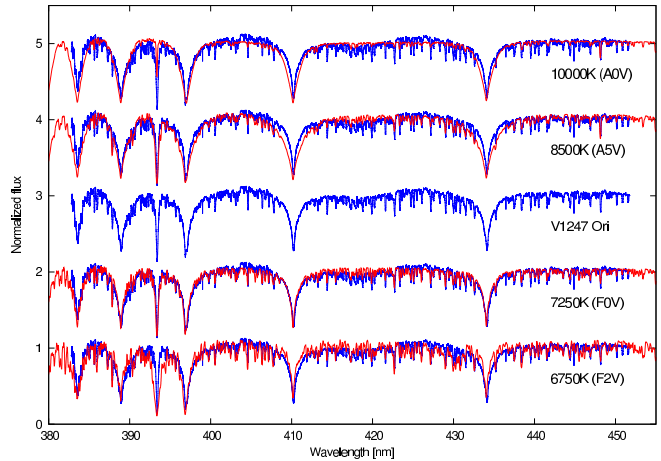}
  \caption{
    Comparison of the HARPS spectrum of V1247\,Ori (3rd from top) 
      for the spectral window between 380 and 460\,nm with 
      synthetic spectra \citep{mun05} for effective temperatures of 10,000\,K (top),
      8500\,K (2nd from top), 7250\,K (4th from top), and 
      6750\,K (5th from top). In order to enable an easy comparison,
      we overplot the synthetic spectra (red curves) with the measured 
      HARPS spectrum (blue curves).
  }
  \label{fig:specclass}
\end{figure}

\begin{figure}[t]
  \centering
  \includegraphics[angle=0,scale=0.8]{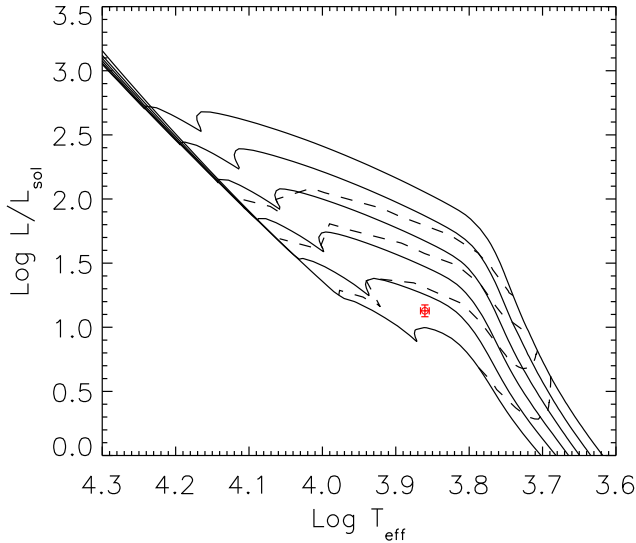}
  \caption{
    HR diagram, showing the location of V1247\,Ori overplotted
    on evolutionary tracks from \citet{bre12}.
    We show a series of isochrones from 1\,Myr to 10\,Myr in steps of 0.2~dex (solid lines, from top to bottom) 
    and evolutionary tracks from 1.5 to 3.0\,M$_{\sun}$ in steps of 0.5\,M$_{\sun}$ (dashed lines).
  }
  \label{fig:hrd}
\end{figure}

\begin{deluxetable}{lccc}
\tabletypesize{\scriptsize}
\tablecolumns{5}
\tablewidth{0pc}
\tablecaption{Stellar parameters of V1247\,Ori\label{tab:resultsspecclass}}
\tablehead{
\colhead{Parameter}      & \colhead{}    & \colhead{Unit}    & \colhead{Value}}
\startdata
Effective temperature    & $T_{\rm eff}$   & K                 & 7250 \\
Gravity                  & $\log g$      & cm\,s$^{-2}$       & 4.5 \\
Metallicity              & [M/H]         &                   & -0.5 \\
Microturbulent velocity  & $\xi$         & km\,s$^{-1}$       & 2 \\
Rotation velocity        & $v_{\rm rot}$   & km\,s$^{-1}$       & 60 \\
Enhancement parameter    & [$\alpha$/Fe] &                   & 0 \\
Distance                 & $d$           & pc                & $385 \pm 15$ \\
Age                      &               & Myr               & $7.4 \pm 0.4$ \\
Mass                     & $M_{\star}$     & $M_{\sun}$         & $1.86 \pm 0.02$ \\
Stellar radius           & $R_{\star}$     & $R_{\sun}$         & $2.3 \pm 0.3$ \\
Radial velocity          & $V_{\rm rad}$   & km\,s$^{-1}$       & $+20.7 \pm 0.5$
\enddata
\tablecomments{For details about the individual parameters and how they have
been derived, we refer the reader to Sect.~\ref{sec:resultsspecclass}.
The radial velocity has been converted to the heliocentric system.}
\end{deluxetable}

In the literature, an unusually wide range of spectral types has been assigned
to V1247\,Ori, ranging from F0V \citep{vie03} to A5III \citep{sch71}.
This has motivated us to re-evaluate the spectral classification by
fitting the HARPS spectrum with the synthetic spectra in the library of 
\citet{mun05},
which covers photospheres over a wide range of values for
effective temperature $T_{\rm eff}$, gravity $\log g$,
metallicity [M/H], microturbulent velocity $\xi$, 
rotation velocity $v_{\rm rot}$, and
[$\alpha$/Fe] enhancement.
For a detailed description of the parameters and
the grid properties, we refer the reader to 
\citet{mun05}.
The range of possible rotation velocities is rather coarsely 
sampled in the original grid.
Therefore, we employed the unbroadened spectra and convolved 
them with the line spread function for a limb-darkened
rotating photosphere. Our new grid samples the rotation velocity
with a resolution of 10\,km\,s$^{-1}$.
We selected the best-fit model based on the $\chi^{2}$-value
and list the best-fit parameters in Tab.~\ref{tab:resultsspecclass}.
These parameters correspond to a spectral type F0V
(based on the table provided by \citealt{kur93}),
which confirms the classification by \citet{vie03}.
Fig.~\ref{fig:specclass} shows the V1247\,Ori HARPS spectrum (blue curve) 
overplotted with the best-fit template spectrum (labeled F0V) 
and template spectra for stars of spectral type 
A0V ($T_{\rm eff}=9520$\,K, $\log g=+4.0$\,cm\,s$^{-2}$),
A5V ($T_{\rm eff}=8200$\,K, $\log g=+4.0$\,cm\,s$^{-2}$), and
F2V ($T_{\rm eff}=6890$\,K, $\log g=+4.0$\,cm\,s$^{-2}$).

As distance for V1247\,Ori we adopt $385 \pm 15$\,pc, 
which is the value determined by \citet{cab08a} and \citet{ter07} 
for other members of the Alnilam cluster.
From the distance and the apparent V-band magnitude 
\citep[$m_{V}=9.861 \pm 0.033$\,mag][]{cab10}, 
we derive the absolute magntiude ($M_{V}=1.934 \pm 0.118$\,mag).
Using the colour excess \citep[E(B-V)=0.02,][]{vie03}, and a bolometric
correction for a F0V star (BC$_{V}=-0.008$; assuming an age of 6.3\,Myr 
and a mass of 1.9\,$M_{\sun}$; \citealt{bre12}),
we estimate the bolometric luminosity to $L_{\rm bol}=13.43 \pm 1.5\,L_{\sun}$.
The radius has then been derived to $R_{\star}=2.3 \pm 0.3\,R_{\sun}$
using the relation $L_{\rm bol}=4 \pi R_{\star}^2 \sigma T_{\rm eff}$, 
where $\sigma$ is the Stefan-Boltzmann constant.

Using the pre-main-sequence evolutionary tracks from \citet[][PARSEC release V1.0]{bre12},
we estimate the mass of V1247\,Ori to $1.86 \pm 0.02$\,M$_{\sun}$,
with an age of $7.4 \pm 0.4$\,Myr (Fig.~\ref{fig:hrd}).
The quoted statistical error bars have been derived assuming the 
aforementioned luminosity uncertainties and an
uncertainty of 100\,K on the effective temperature, but do not
include systematic uncertainties that might be associated with the calibration
of the stellar evolutionary models.
The derived age is consistent with the age of the Alnilam cluster 
\citep[5-10\,Myr,][]{cab08b}.

\subsection{Spectral energy distribution, line emission identification, and variability}
\label{sec:resultsdiskSED}

\begin{figure}[t]
  \centering
  \includegraphics[angle=0,scale=0.5,trim=0mm 0mm 0mm 0mm,clip]{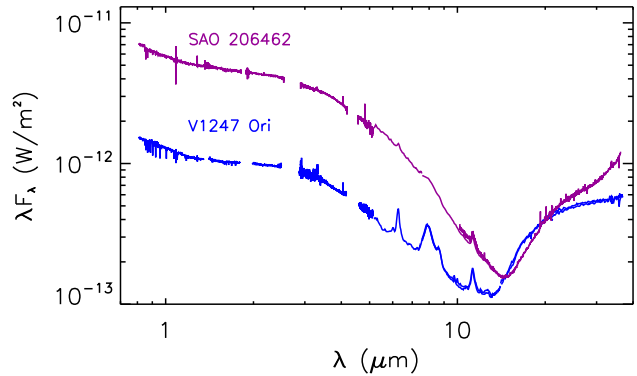} 
  \caption{
    Comparison of the SED of V1247\,Ori with the SED of another
    star with a pre-transitional disk (HD\,135344B; data from 2009 May 20).
  }
  \label{fig:SEDcomparison}
\end{figure}

\begin{figure}[t]
  \centering
  \includegraphics[angle=270,scale=0.35]{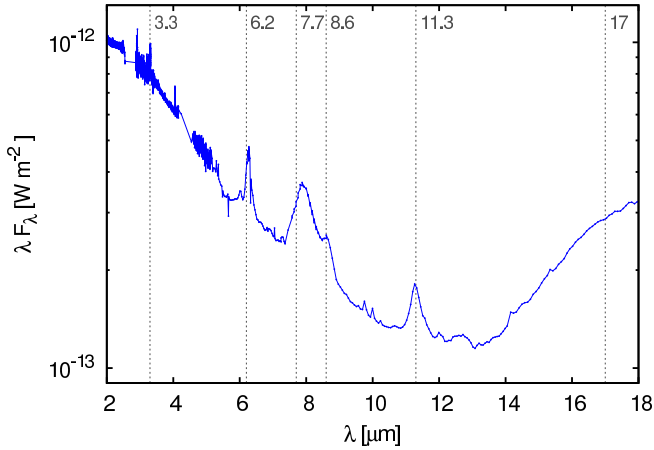}  
  \caption{
    V1247\,Ori spectrum between 2 and 18\,$\mu$m, showing 
    strong line emission, where the strongest lines appear to be associated with
    hydrocarbon features (grey dotted lines).
  }
  \label{fig:PAH}
\end{figure}

\begin{figure}[t]
  \centering
  \includegraphics[angle=270,scale=0.35]{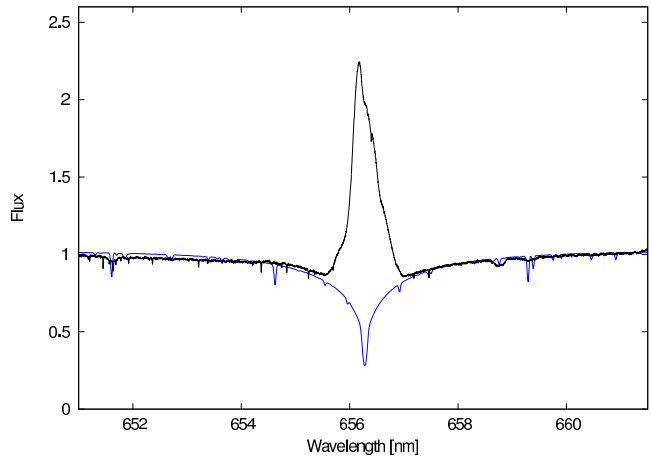}  
  \caption{
    HARPS spectrum of V1247\,Ori around the H$\alpha$-line (black line, original resolution),
    overplotted with a template spectrum for a F0V photosphere (blue  line).
    The spectrum has not been corrected for telluric features.
  }
  \label{fig:HARPSHalpha}
\end{figure}

\begin{figure}[t]
  \centering
  \includegraphics[angle=0,scale=0.555,trim=0mm 0mm 0mm 8mm,clip]{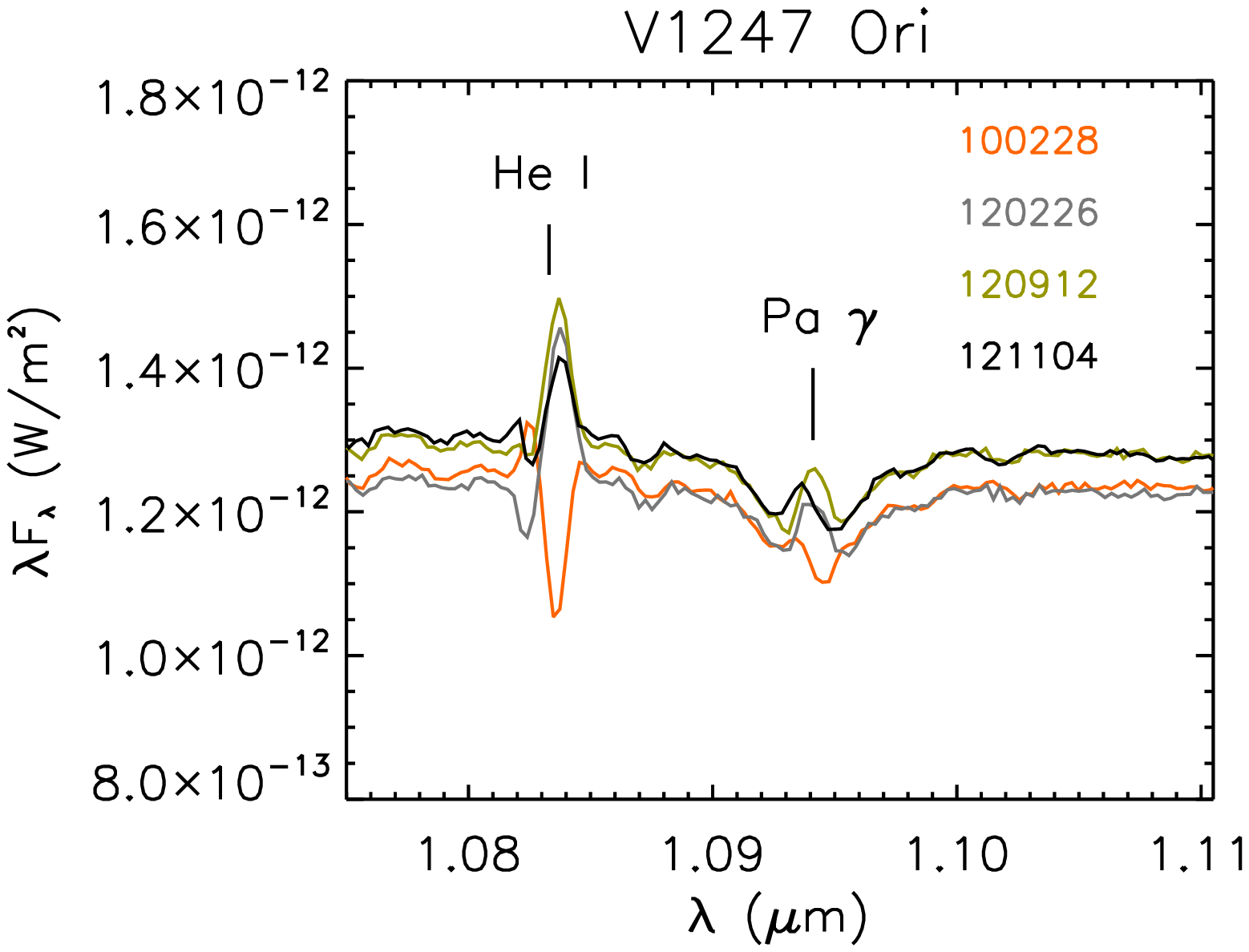}\\
  \includegraphics[angle=0,scale=0.36,trim=0mm 0mm 0mm 13mm,clip]{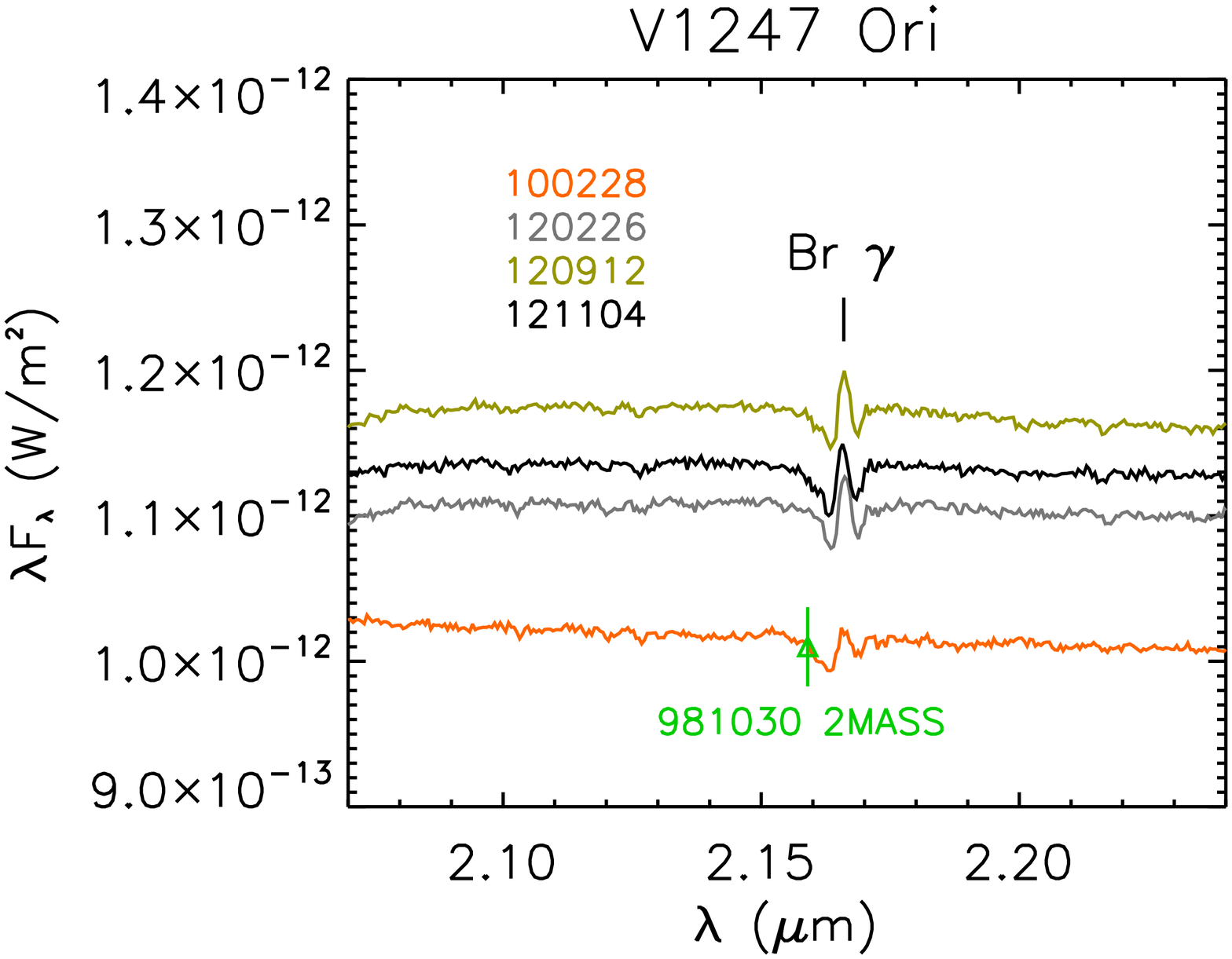}  
  \caption{
    SpeX spectra covering the \ion{He}{1} and Pa$\gamma$ line (top panel) as well as the 
    Br$\gamma$ line (bottom panel) at four epochs.
  }
  \label{fig:SPEXBrGHePa}
\end{figure}

\begin{figure}[t]
  \centering
  \includegraphics[angle=0,scale=0.52,trim=0mm 0mm 0mm 8mm,clip]{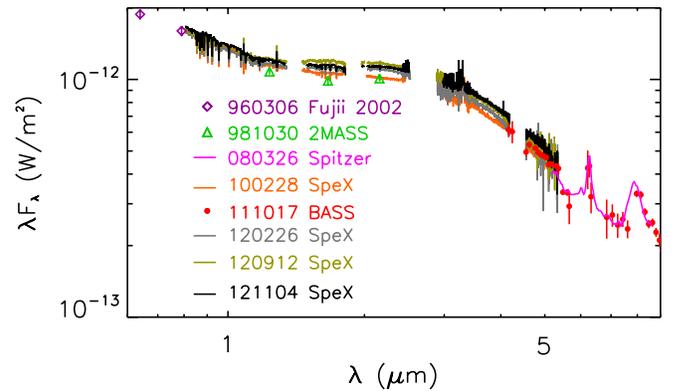}   
  \caption{
    NIR/MIR spectra (2010 February to 2012 November) and 
    photometry data (1996 March to 2008 March 26)
    revealing significant variability on time scales of several
    months to years.
  }
  \label{fig:SPEXspectrumvar}
\end{figure}

The SED of V1247\,Ori shows remarkable similarities with the SED of other
stars with pre-transitional disks, such as \object{HD\,135344B}
\citep[=SAO\,206462,][]{gra09,sit12} or \object{HD\,169142} \citep{gra07,pan08,hon12},
including a strong infrared excess emission with a significant
flux depression in the $\sim$3--15\,$\mu$m mid-infrared wavelength regime 
compared to classical T~Tauri disks (Fig.~\ref{fig:SEDcomparison}).
These spectral characteristics are often interpreted as evidence for
a gapped disk structure, where the near-infrared ($<3\,\mu$m) emission
is attributed to coming from a hot {\it inner} disk 
and the mid-infrared ($>10\,\mu$m) emission to colder material
in the {\it outer} disk.

In the $K$ band, no lines corresponding to the CO overtone ($\Delta v=2$, around 2.3\,$\mu$m)
or \ion{He}{1} (2.058\,$\mu$m) have been detected.
However, our spectra show some extremely strong emission line components
around 3.3, 6.2, 7.7, 8.6, and 11.3\,$\mu$m (Fig.~\ref{fig:PAH}), which we identify 
as hydrocarbon-related features (sometimes incorrectly referred to in the literature as 
polycyclic aromatic hydrocarbon (PAH) features, e.g.\ \citealt{van04c}).
We also detect accretion-tracing spectral lines,
including H$\alpha$ (Fig.~\ref{fig:HARPSHalpha}), 
\ion{He}{1}, Pa$\gamma$, and Br$\gamma$ (Fig.~\ref{fig:SPEXBrGHePa}),
which are superposed on the underlying photospheric component.
The equivalent width of H$\alpha$ is -0.295\,nm and
-1.006\,nm after we correct for the strong underlying photospheric absorption 
component (Fig.~\ref{fig:HARPSHalpha}).
The H$\alpha$-line profile is slightly asymmetric with a weak redshifted
absorption component, showing a similar morphology as predicted
by magnetospheric accretion models for disks seens closer to edge-on than face-on \citep{muz98c}.

The Br$\gamma$ line is rather weak, but shows significant variability,
which is most evident between our epoch in 2010 and the 
three epochs in 2012 (Fig.~\ref{fig:SPEXBrGHePa}).
This variability in Br$\gamma$ seems to be
correlated with the Pa$\gamma$ and \ion{He}{1}-line luminosity (Fig.~\ref{fig:SPEXBrGHePa}, top),
where \ion{He}{1} even switches from absorption to emission between
2010 February 28 and 2012 February 26.
Furthermore, there appears to be a correlation between the
equivalent width of the Br$\gamma$ line and
NIR continuum infrared excess (Fig.~\ref{fig:SPEXBrGHePa}, bottom).

Our multi-epoch spectro-photometric data from visual to MIR wavelengths are overplotted
in Fig.~\ref{fig:SPEXspectrumvar} (and in Fig.~\ref{fig:MIDIspectrum} for our MIDI and BASS spectra), 
revealing significant variability, which seems to have the strongest amplitude 
(15\%) in the $K$ band, with a calibration uncertainty of 5\%.
Given that the amplitude of the continuum variability seems to peak in the $K$ band and decreases 
both towards shorter and longer wavelength, we consider it most likely that the origin of the variability
is associated with the inner disk, possibly with rotation-induced projection effects, 
as discussed in Sect.~\ref{sec:discussion}.
Photometric and/or spectroscopic variability has also been found for other (pre-)transitional disks 
\citep[e.g.][]{muz09,esp11,sit12,fla12}.

\subsection{Spatially resolved constraints on the gapped disk structure}
\label{sec:resultsdisk}

\subsubsection{Geometry of the outer disk and gap region}
\label{sec:resultsdiskMIR}

\begin{deluxetable}{cccccc}
\tabletypesize{\scriptsize}
\tablecolumns{6}
\tablewidth{0pc}
\tablecaption{Results from our geometric model fits
  to the T-ReCS data\label{tab:modelfitTReCS}}
\tablehead{
\colhead{Geometry} & \colhead{$\lambda_{c}$} & \colhead{FWHM}  & \colhead{$R_{\rm ring}$} & \colhead{$i$}       & \colhead{PA} \\
                                   & \colhead{[$\mu$m]}     &  [mas]                     & \colhead{[mas]}             & \colhead{[\deg]}  & \colhead{[\deg]} }
\startdata
GAUSS         & 8.8       & $113 \pm 17$ &                         & $40.8 \pm 2$            & $119 \pm 9$ \\
GAUSS         & 9.7       & $135 \pm 8$   &                          & $31.8 \pm 7$           & $104 \pm 15$ \\
GAUSS         & 11.7     & $130 \pm 4$   &                          & $31.5 \pm 6.5$       & $113 \pm 12$ \\
RING            & 8.8       &                           & $64 \pm 9$ & $38.2 \pm 2.5$      & $119 \pm 9$ \\
RING            & 9.7       &                           & $79 \pm 5$ & $31.3 \pm 7.5$      & $104 \pm 15$ \\
RING            & 11.7     &                           & $75 \pm 1$   & $30.7 \pm 6.3$      & $114 \pm 13$
\enddata
\end{deluxetable}

Our T-ReCS speckle observations resolve the disk structure at 
mid-infrared wavelengths of 8.7, 9.7, and 11.7\,$\mu$m.
Providing a complete two-dimensional $uv$-plane coverage to baseline lengths of
$5.5$\,m, these data are perfectly suited to constrain the geometry of
the large-scale disk and to determine the disk orientation 
and inclination angle.
A visual inspection of the six independent 
power spectra (Fig.~\ref{fig:TReCSpowerspec}) reveals a significant elongation along PA$\sim 20^{\circ}$ 
(corresponding to an orientation of the disk major axis along PA$\sim 110^{\circ}$).
In the power spectra for the Si-3 filter, this elongation pattern is superposed with another weaker signature,
resulting in an approximately cross-shaped power spectrum
(Fig.~\ref{fig:TReCSpowerspec}, middle row).
Given that this pattern appears only the in Si-3 data set,
which has the lowest SNR due to the atmospheric ozone absorption,
we consider it likely that this signature is caused by
instrumental artefacts, such as wind-induced vibration in the telescope.
In the probed high-visibility regime ($V^{2} \gtrsim 0.5$), the visibility profile 
is rather insensitive to the detailed brightness distribution,
which allows us to derive the disk orientation and inclination 
using a simple elliptical Gaussian model,
without taking the inner disk structure into account.
The models include a photospheric flux component, although the 
contributions from this spatially unresolved component
are practically neglible ($<2$\% based on our SED decomposition).

The aim of our Gaussian model fitting is to
quantify the elongation that is present in the
2-D power spectra (Fig.~\ref{fig:TReCSpowerspec}).
Using a Levenberg-Marquardt $\chi^{2}$-minimization
algorithm, we determine the best-fit 
full width half maximum (FWHM), the position angle of the 
major axis of the elliptical Gaussian (PA), and the inclination angle ($i$).
We performed the fit for each of our wavelength bands independently and 
we estimate the parameter uncertainties for each wavelength band from the 
scatter between our two independent measurements.
The determined parameters are listed in Tabble~\ref{tab:modelfitTReCS}
and consistently show an elongation along position angle 
PA=$110^{\circ}$ with an inclination angle $i=35^{\circ}$ (i.e.\ closer to face-on).
The Gaussian model provides a very simple and robust estimation for
the elongation of the MIR emission component, but does
not take the additional information provided by the MIDI data
into account, which reveals the presence of a compact spatial component.

The choice for the second model geometry (elliptical ring + 
unresolved component) is motivated by our combined MIDI+T-ReCS data set,
which shows that the visibility function at wavelengths of 8.7\,$\mu$m 
drops significantly at short baselines ($V^2=0.5$ at $< 5.5$\,m), but
then stays rather constant at a level of $V^2 \approx 0.1$ from 46\,m
to 125\,m (Fig.~\ref{fig:visprofile-innerouter}, 3rd panel from top).  
It is important to note that this visibility profile is 
substantially different from the ones observed towards 
classical T~Tauri or Herbig~Ae/Be stars \citep[e.g.][]{kra08a}
and provides model-independent evidence for the presence of 
two spatial components, namely an extended component
(which we represent with a ring geometry)
and a compact component that is only marginally 
resolved even at hectometric baseline lengths.
Denoting the flux contributions of the ring component with 
$F_{\rm  ring}$, the flux of the unresolved component with 
$F_{\rm  unres}$, the total measured visibility will be given by
$V_{\rm total}(B) = (F_{\rm ring} V_{\rm ring}(B) + F_{\rm unres}) / 
(F_{\rm ring} + F_{\rm unres})$, where $V_{\rm ring}(B)$ is the
visibility profile of the ring component.
At long baselines (where the extended component is overresolved,
i.e. $V_{\rm ring} \approx 0$), $V_{\rm total}(B)$ will converge
towards $F_{\rm unres} / (F_{\rm ring} + F_{\rm unres})$.
According to our MIDI measurements, this convergence value 
is $\sim 0.3$, which leads us to fix 
$F_{\rm unres} / (F_{\rm ring} + F_{\rm unres}) = 0.3$ in our model.
The fractional width of the ring has been fixed to $0.2$.
The best-fit parameters are listed in Tab.~\ref{tab:modelfitTReCS},
where we denote the best-fit ring radius with $R_{\rm ring}$.

This two-component model yields our best estimate
for the disk inclination angle ($i = 31.3 \pm 7.5^{\circ}$) and
disk position angle (PA=$104 \pm 15^{\circ}$).
The existing Keck and VLTI long-baseline interferometric 
observations are not able to further constrain these parameters,
but provide highly complementary constraints on the
detailed object geometry, as will be discussed in the following sections.

\subsubsection{Geometry of the inner disk}
\label{sec:resultsdiskNIR}

Our AMBER+ASTRA interferometric observations trace a compact emission 
component with a characteristic size of $0.90\pm0.38$\,mas at 2\,$\mu$m (ring radius),
assuming a ring with fraction width of 20\%, as adopted, for instance, by \citealt{mon02}.
This is more than 80 times smaller than the extended MIR component seen by 
our MIDI+T-ReCS MIR observations (79\,mas at 12\,$\mu$m).
With measured squared visibilities down to $V^{2}=0.6$, the emission is clearly resolved 
both in the $H$ and $K$ bands.
The visibility level drops slightly towards shorter wavelengths 
(Fig.~\ref{fig:VISAMBER}, second panel from top).
However, when plotted against spatial frequency (Fig.~\ref{fig:visprofile-innerouter}, 2nd panel from top),
the visibilities at all wavelengths follow the same visibility function.
This indicates that the visibility drop simply reflects the increase in angular resolution towards 
shorter wavelengths, but not a temperature gradient in the emitting structure,
nor the presence of an optically thick inner gaseous emission component,
such as observed in the disks around some Herbig~Ae/Be stars \citep[][]{eis07a,kra08a,ise08}.
We also do not find any indications for an extended NIR scattered-light halo, 
such as observed for Herbig~Ae/Be stars like R\,CrA \citep{kra09b}.

Unfortunately, the current $uv$-coverage of the AMBER+ASTRA observations is 
not sufficient to measure the inclination angle of the inner disk independently.
Therefore, we assume in our following modeling that the inner disk is coplanar
to the outer disk.

The characteristic size of the NIR-emitting region (ring radius $0.90\pm0.38$\,mas or $0.34\pm 0.14$\,AU at $d=385\pm 15$\,pc)
is significantly larger than the expected size of the dust sublimation radius (0.085\,AU)
as computed for black dust properties (cooling efficiency $\epsilon = 1$; sublimation temperature 1500\,K)
using Eq.~12 in \citet{dul10}, which includes back-warming effects.
In Sect.~\ref{sec:modeling-optthin}, we suggest that this increased NIR size 
is due to contributions from optically thin material located at larger stellocentric radii.
Alternatively, it would be possible to reconcile the measured and expected sizes by
assuming a population of smaller dust grains with cooling efficiency $\epsilon \sim 0.2$.

The closure phases measured with AMBER are zero within the
uncertainties of $\sim 5^{\circ}$ (Fig.~\ref{fig:VISAMBER}), 
which is consistent with a centro-symmetric brightness distribution
on scales of $\rho \lesssim 30$\,mas ($r \lesssim 12$\,AU).

\subsubsection{Asymmetric structures in the gap region}
\label{sec:resultsNIRC2}

\begin{deluxetable}{ccccc}
\tabletypesize{\scriptsize}
\tablecolumns{4}
\tablewidth{0pc}
\tablecaption{Results from the companion model fit to the NIRC2 data\label{tab:modelfitNIRC}}
\tablehead{
\colhead{Spectral} & \colhead{$\Delta m$}  & \colhead{$\rho$}       & \colhead{PA} \\
\colhead{band}      &  \colhead{[mag]}                               &  \colhead{[mas]}                         & \colhead{[\deg]}  }
\startdata
$H$         & $5.73 \pm 0.31$ & $94.9 \pm 3.0$   & $61.9 \pm 1.9$  \\
$K'$        & $5.22 \pm 0.18$ & $42.8 \pm 5.4$   & $306.6 \pm 3.4$ \\
$L'$        & $6.05 \pm 0.35$ & $114.4 \pm 9.2$ & $328.6 \pm 4.4$
\enddata
\end{deluxetable}

Measuring closure phases with sub-degree accuracy, 
our NIRC2 aperture masking observations are sensitive to 
asymmetries in the brightness distribution on angular scales well 
below the diffraction limit of the 10\,m Keck telescope.
The data reveal non-zero phases in all three wavelength bands, 
with varying significance levels.
In order to explore the origin of these asymmetries, we fit a star+companion 
model to our data, where we treat the three wavelength filters separately.
The error estimation and detection limits in our code have been
carefully tested and fine-tuned in various earlier studies \citep[e.g.][]{kra11,kra12d}.
The fitting results are summarized in Tab.~\ref{tab:modelfitNIRC}, while
the corresponding best-fit phase plots and $\chi^2$-maps are shown in Fig.~\ref{fig:phases-NIRC2}.

The $K'$-band data provides convincing evidence for the detection of a companion
at a separation of {$0.043 \pm 0.005$\arcsec} from the star and towards PA=$307\pm 3^{\circ}$.
With more than $10{\sigma}$, the detection is comparable to the significance level of other
companion detections around transitional or pre-transitional disks with the aperture masking technique.
From the brightness ratio ($\Delta K=5.22 \pm 0.18$\,mag) and the 2MASS $K$-band magnitude 
($m_{K,\star}=7.41 \pm 0.03$\,mag), we estimate the absolute magnitude of this potential companion to 
$M_{K} = 4.70 \pm 0.29$\,mag for a distance of $385\pm 15$\,pc.
Based on the evolutionary models from \citet{bar98}, we estimate
the mass to $M_{\mathrm{comp}}=0.3\,M_{\sun}$ at 6\,Myr.

Fitting our companion model to the $H$ and $L'$-band data
yields detections at $\sim 3\sigma$ or more.
However, for the following reasons, these detections are in conflict with the
companion interpretation of the $K'$-band detection:
\begin{enumerate}
\item The companion positions in the $H$ and $L'$-band best-fit model
  deviate significantly from the position derived from the $K'$-band data (Tab.~\ref{tab:modelfitNIRC}).
\item At the location of the $K'$-band detection,
  we obtain $3\sigma$ non-detections with magnitude limits of $\Delta H_{\rm lim}=5.7$\,mag and $\Delta L'_{\rm lim}=5.6$\,mag.
  These upper limits place strong constraint on the color of the putative companion,
  yielding an unrealistic red color of $\Delta (H-K) \ge 0.48$\,mag compared to the star+inner disk system.
  For a $0.3\,M_{\sun}$ companion, we would expect to detect the companion with a color of $\Delta (H-K)=0.26$\,mag.
  Therefore, in order to reconcile the companion scenario with the color constraints, we would have to assume
  that the color of the companion is reddened by inner disk shadowing and/or that the phase signal in the other
  bands is reduced by contamination due to disk emission.
\item For the $L'$-band data, we find several solutions (Fig.~\ref{fig:phases-NIRC2}, bottom right), 
  indicating that the companion model is not a good representation
  of the true object morphology.
\end{enumerate}
We interpret this as strong evidence that the measured asymmetries 
{\it do not} trace a single point-source, but distinct asymmetric structures on 
different spatial scales.
The model fits suggest that these asymmetric structures are located on spatial scales 
of $\sim$ 40--110\,mas (Tab.~\ref{tab:modelfitNIRC}) or $\sim$ 15--40\,AU,
i.e.\ within the gap region.
This interpretation is in line with our disk model (Sect.~\ref{sec:modeling-optthin}), 
which predicts that the optically thin gap material contributes a significant fraction
of the $H$-, $K'$-, and $L'$-band emission (Fig.~\ref{fig:visprofile-optthin}, top panel).
Any inhomogenities in the distribution of this optically thin gap material introduces asymmetries
that appear as a non-zero phase signal in aperture masking data, mimicking a companion signal.
For complex disk structures, such as spiral arms, the amplitude and direction
of the asymmetry changes with wavelength, as different wavelengths are sensitive 
to emission from different stellocentric radii.
The non-zero phases measured by our aperture masking data 
likely trace such asymmetric structures in the disk gap.
Given its limited angular resolution, our aperture masking data is not able 
to constrain the detailed spatial structure of these asymmetric features,
although this should become feasible with future long-baseline 
interferometric aperture synthesis imaging observations.

\subsection{Multi-wavelength interferometry + SED modeling of the disk structure}
\label{sec:modeling}

In order to constrain the disk geometry quantitatively, we combine in this section
our SED and NIR+MIR interferometric data and fit it using a model of a 
geometrically thin dust disk.
Our model assumes that the intensity $I_{\nu}$ at each surface element in the disk
can be described as
\begin{equation}
  I_{\nu} = B_{\nu}(T(r)) (1-e^{-\tau_{\nu}}),
\end{equation}
where $B_{\nu}$ is the Planck function and $T(r)$ is the temperature at the considered disk radius,
and $\tau_{\nu}$ is the optical depth, which is related to the surface density $\Sigma$ 
and the dust opacity $\kappa_{\nu}$ at frequency $\nu$ with $\tau_{\nu} = \kappa_{\nu} \Sigma$. 
The radial dust temperature is 
computed based on the thermal balance between stellar heating
and radiative cooling 
using the relation presented by \citet[][]{dul10}.
This approach provides a first-order approximation of the intensity profile, sufficient for our
purposes, 
but does not take the vertical disk structure, nor shadowing effects 
\citep[e.g.][]{esp10} into account.  These effects should be constrained
in future studies using a more sophisticated multi-dimensional radiative transfer 
and hydrodynamics simulation approach.
Assuming a point-symmetric brightness distribution, 
the model is also not able to reproduce the asymmetries that we have detected 
with our aperture masking observations (Sect.~\ref{sec:resultsNIRC2}).
For the dust properties we use opacities for silicate
and carbon (graphite) grains by \citet{dra84} and mixtures of these grain species.
We adopt the grain size distribution from \citet{mat77}, 
but tested also grains with a single grain size.

Given our earlier, model-independent indications for an inner and outer disk component (Sect.~\ref{sec:resultsdiskNIR}),
we divide the disk into three regions, namely the 
inner disk (ID, extending from $R_{\rm ID,in} \leq r \leq R_{\rm ID,out}$),
the gap region ($R_{\rm ID, out} < r < R_{\rm OD,in}$), and 
the outer disk (OD, extending from $R_{\rm OD,in} \leq r \leq R_{\rm OD,out}$),
where $R_{\rm ID,in}$, $R_{\rm ID,out}$, $R_{\rm OD,in}$, and $R_{\rm OD,out}$
are treated as free parameters.
For the inner and outer disk, we fix the surface density to
an (arbitrary) very high value, ensuring that the emission is optically thick.
The surface density  in the gap region $\Sigma_{\rm gap}$ is treated as a free parameter
with a flat surface density profile ($\Sigma(r)$ = const.), as motivated by
hydrodynamic simulations of gaps in protoplanetary disks \citep[e.g.][]{kle99,zhu11}.
The model is fitted simultaneously to the SED and the interferometric data and the 
five aforementioned parameters are fitted using a
least-square minimization process.  In order to avoid local minima, we vary the 
initial parameters systematically on a grid.
Given the heterogeneous sampling of the SED, we binned the spectroscopic and
photometric data on an equidistant grid in logarithmic wavelength space.
In the fitting process, the VLTI/AMBER, KI/ASTRA, Keck-II/NIRC2, VLT/NIRC2, VLTI/MIDI, 
and Gemini/T-ReCS data are attributed the same weight.

In addition, our model includes the stellar photosphere,
which we represent with the recommended SYNTHE model 
for a F0V star \citep[$\log g=4.5$, $T_{\rm eff}=7250$\,K,][]{kur93}
We compute the flux ratio between the photospheric
and circumstellar emission for each wavelength channel separately,
yielding values from 45\% at $1.5\,\mu$m, 
21\% at 1.9\,$\mu$m, to 12\% at 2.5\,$\mu$m.

\subsubsection{Inner \& outer disk (ID+OD model)}
\label{sec:modeling-innerouter}

\begin{figure}[htbp]
  \centering
  \includegraphics[angle=270,scale=0.29]{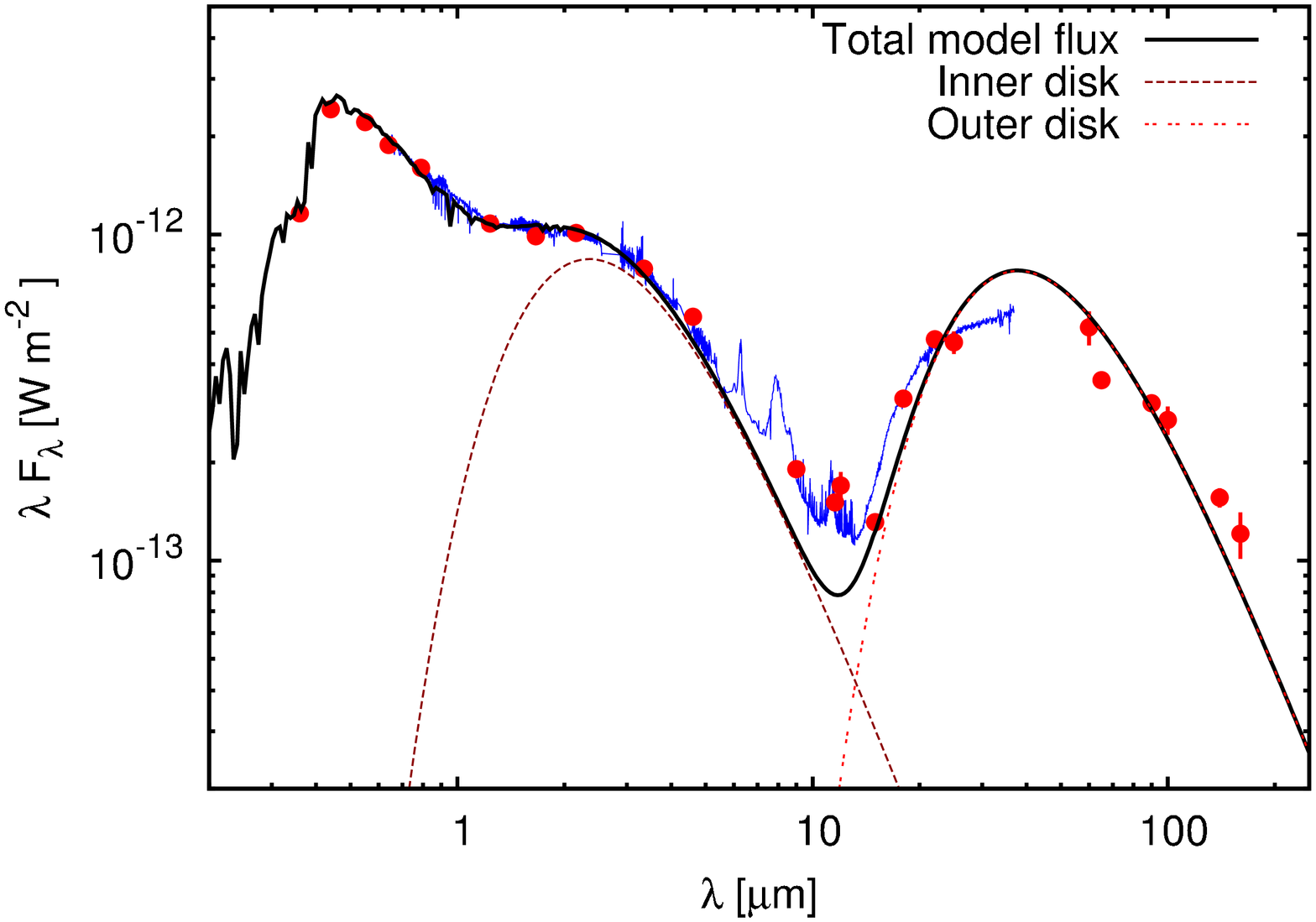}\\
  \includegraphics[angle=0,scale=0.54]{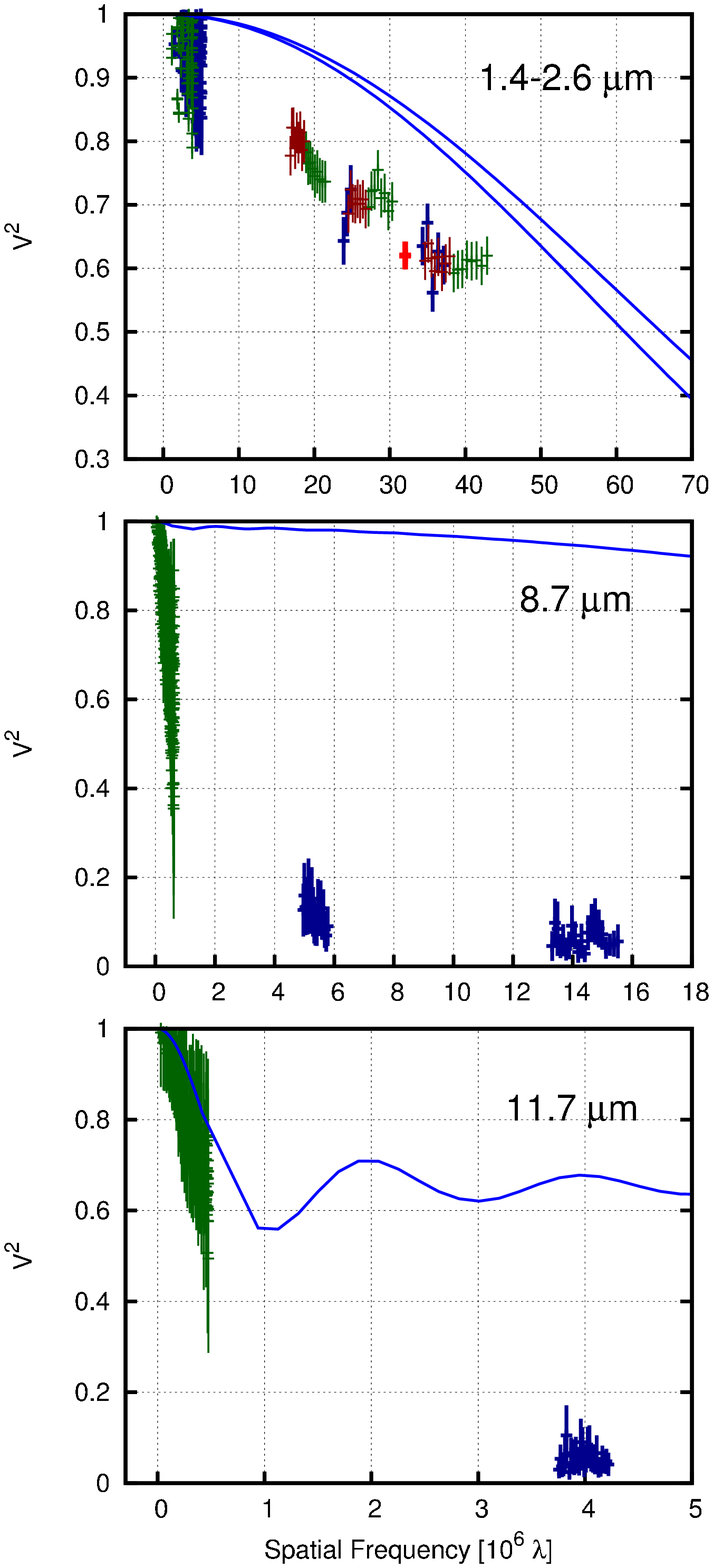}      
  \caption{
    SED (top panel) and squared visibility amplitudes (2nd to 4th panel), compared with the
    inner \& outer disk model (``ID+OD'' model, Sect.~\ref{sec:modeling-innerouter}).
    The data points at NIR wavelengths are color-coded as follows:
    blue: VLTI/AMBER+Keck/NIRC2, $H$ band; 
    red: KI/V2-SPR, $K$ band; dark-red: VLTI/AMBER, upper $K$ band;
    green: VLTI/AMBER+Keck/NIRC2, lower $K$ band.
    The 2nd panel covers the wavelength range 1.4-2.6\,$\mu$m ($H$ and $K$ band)
    and the shown two curves cover the extreme values of this wavelength range.
  }
  \label{fig:visprofile-innerouter}
\end{figure}

A visual inspection of the SED (Fig.~\ref{fig:SED}) suggests that
two simple blackbody components might be sufficient to fit the full SED.
Therefore, in a first modeling attempt, we include only two optically
thick emission components, without emission from inside the gap ($\Sigma_{\rm gap} = 0$).
A standard dust mixture of 50\% silicate + 50\% carbon was used for this model.

The resulting best-fit is shown in Fig.~\ref{fig:visprofile-innerouter}
and can reproduce the 
SED at NIR and far-infrared wavelengths
reasonably well, even though it underestimates the flux in the MIR regime.
The corresponding parameters are listed in Tab.~\ref{tab:modelfit}.

We find that this model results in much too high MIR visibilities, indicating that the
MIR-emitting structure in the model is too compact.
Improving the fit cannot be achieved by extending the inner disk ($R_{\rm ID, in}$),
since this would result in a conflict with the NIR visibilities.
Increasing $R_{\rm ID,out}$,  on the other hand, changes the shape
of the NIR excess and leads to an inconsistent MIR SED,
suggesting that the optically thick inner disk is very narrow.

The main inadequacy of the ID+OD model is its failure to reproduce the
measured low MIR visibilities, indicating that this model lacks an emission 
component on spatial scales at least one order of magnitude larger than 
the inner disk.  
Therefore, we investigated whether parts of the outer disk might be able 
to contribute extended MIR emission, for instance from a vertically 
extended wall located at the inner truncation radius of the outer disk.
In order to reproduce the MIR visibilities, this emission has to
originate from stellocentric radii $r \gtrsim 46$\,AU, while temperatures
$T \gtrsim 400$\,K are required in order to contribute significant
amounts of thermal emission in the 8 to 13\,$\mu$m wavelength regime.
However, we were not able to find a physical scenario that would
allow us to achieve such high temperatures at the required stellocentric
radii of $r \gtrsim 46$\,AU.  For black dust properties ($\epsilon=1$), 
the expected temperatures at 46\,AU are $T \sim 80$\,K.
This conflict cannot be resolved by adopting small grain species, such as 
$0.1\,\mu$m-sized carbon grains ($\epsilon=0.12$), where the 
temperature would still be only $T \sim 130$\,K.
Therefore, we conclude that the discrepancies of the ID+OD model
cannot be improved by changing the structure of the outer disk.
This conclusion is also consistent with the finding from 
earlier SED-based studies \citep[e.g.][]{dal05,esp10}
that investigated the detailed vertical structure of the outer disk wall
of stars in Taurus and Ophiuchus
and found that its contribution shortwards of 20\,$\mu$m are negligle.

\subsubsection{Inner \& outer disk with optically thin gap material (ID+GapMaterial+OD model)}
\label{sec:modeling-optthin}

\begin{figure}[htbp]
  \centering
  \includegraphics[angle=270,scale=0.29]{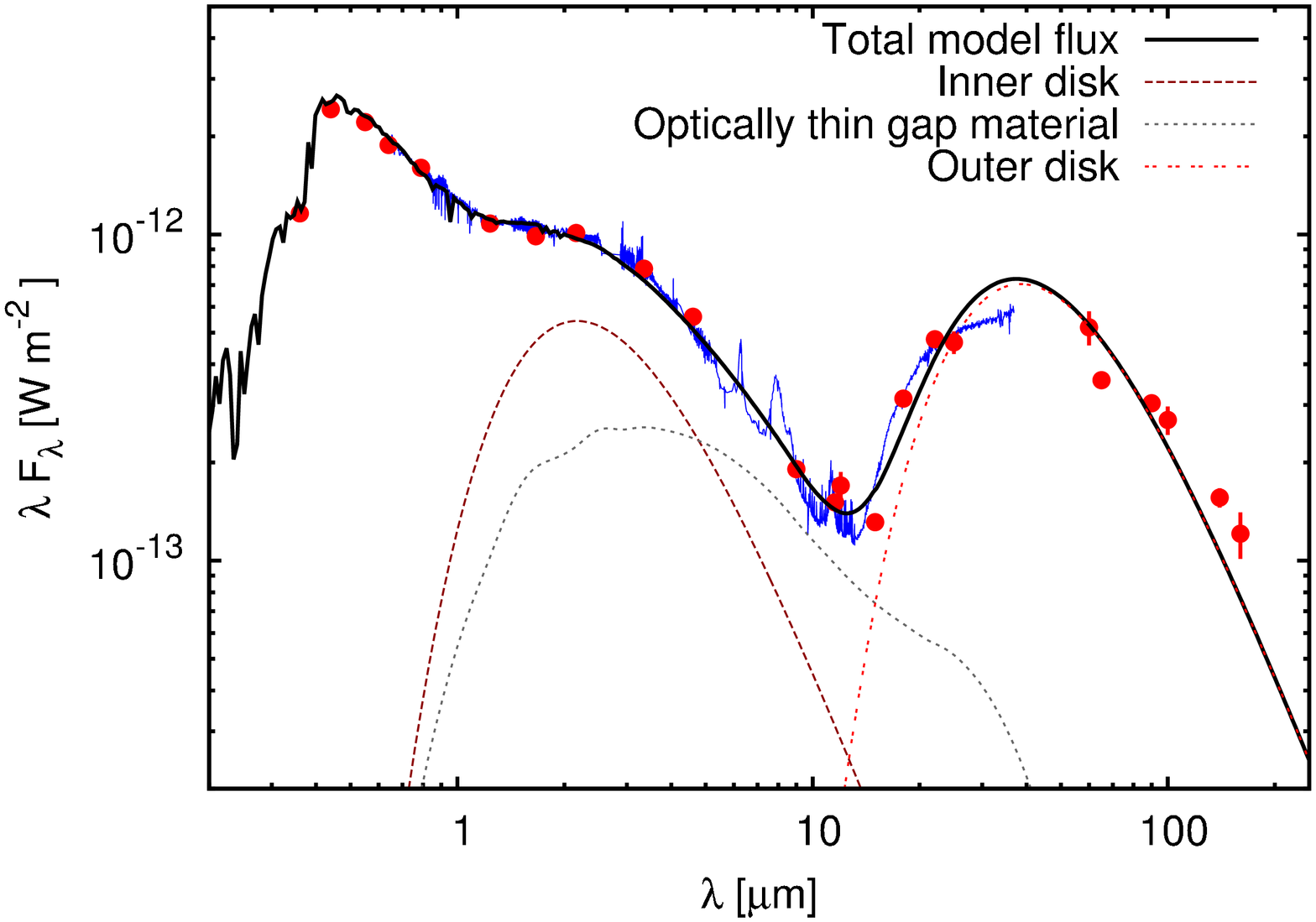}\\
  \includegraphics[angle=0,scale=0.54]{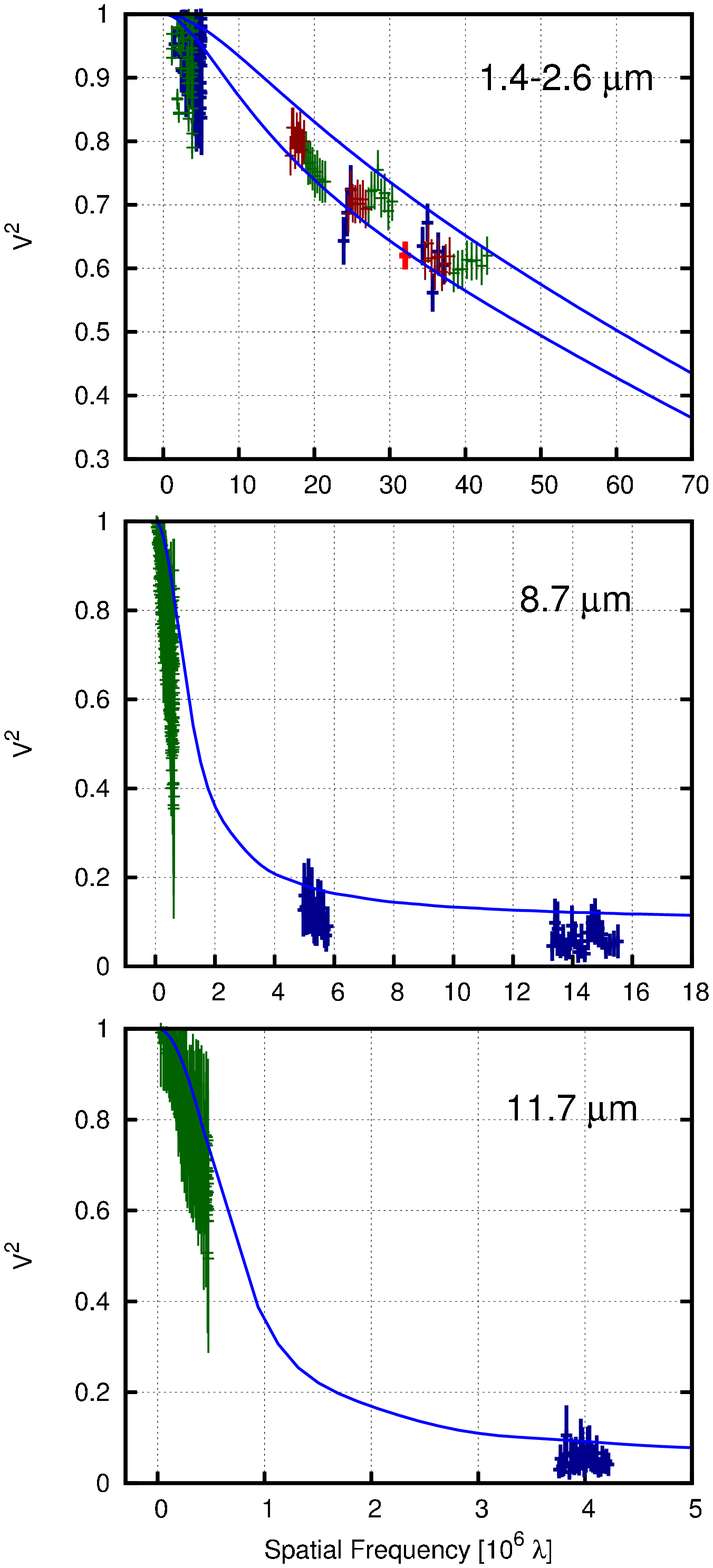}      
  \caption{
    Same as Figure~\ref{fig:visprofile-innerouter}, but for the
    model with optically thin material located in the gap 
    (``ID+GapMaterial+OD'' model; Sect.~\ref{sec:modeling-optthin}).
  }
  \label{fig:visprofile-optthin}
\end{figure}

\begin{deluxetable*}{llccc}
\tabletypesize{\scriptsize}
\tablecolumns{5}
\tablewidth{0pc}
\tablecaption{Best-fit parameters for models described in Sect.~\ref{sec:modeling}
  \label{tab:modelfit}}
\tablehead{
\colhead{Parameter} &  &                              & \colhead{ID+OD} & \colhead{ID+GapMaterial+OD}}
\startdata
Inner Disk, inner radius &  $R_{\rm ID,in}$     & [AU]      & 0.19 & 0.18 \\
Inner Disk, outer radius & $R_{\rm ID,out}$    & [AU]      & 0.34 &  0.27 \\ 
Outer Disk, inner radius &  $R_{\rm OD,in}$    & [AU]      & 44 & 46 \\ 
Outer Disk, outer radius & $R_{\rm OD,out}$   & [AU]      & 85 &  85 \\ 
Dust surface density in gap  & $\Sigma_{\rm gap}$       & [g$\cdot$cm$^{-2}$] & 0 &   $9 \times 10^{-6}$ \\
\hline 
                              & $\chi^2_{r,V}$            &                  & 73.5 &  3.0 \\
                              & $\chi^2_{r,\mathrm{SED}}$    &                  & 3.8 &  1.7 \\
                              & $\chi^2_{r,\mathrm{total}}$  &                  & 65.6 &  2.8
\enddata
\tablecomments{$\chi^2_{r,V}$ denotes the reduced $\chi^2$ likelihood estimator for the visibility data, as defined in \citet[][equation 1]{kra09b}. $\chi^2_{r,\mathrm{SED}}$ is defined equivalent for the SED data and $\chi^2_{r,\mathrm{total}}$ for the fit to the combined visibility and SED dataset.}
\end{deluxetable*}

As a second scenario, we consider that the gap might not be completely depleted of material, 
but filled with optically thin dust.  
For this purpose, we vary both the surface density $\Sigma_{\rm gap}$ 
and the dust properties.

Our best-fit model is shown in Fig.~\ref{fig:visprofile-optthin} 
and includes carbon dust grains \citep{dra84} with a MRN-type
grain size distribution \citep{mat77}.
The dust mineralogy is constrained both by the absence of a 10\,$\mu$m 
silicate feature and the shape of the SED, in particular the very sharp flux
minimum around $\lambda \sim 15$\,$\mu$m (Fig.~\ref{fig:SED}).
We find that small silicate grains are not compatible 
with our observations, since they result in a strong silicate feature.
Large silicate grains, on the other hand, result in too much excess emission 
at wavelengths $\lambda \gtrsim 8$\,$\mu$m.
A carbon dust composition with an MRN grain size distribution
allows us to achieve a good fit with a $\chi^2_{r,\mathrm{total}}=2.8$ (Table~\ref{tab:modelfit}).

Compared to the ID+OD model, the NIR emission contains here
contributions both from the inner disk and from the optically thin
dust in the gap region, which results in lower visibilities and, at the
same time, a more shallow visibility profile, providing a significantly
better fit to the measured $H$ and $K$-band visibilities 
(second panel from top in Fig.~\ref{fig:visprofile-innerouter} and \ref{fig:visprofile-optthin}).   
The improvement is even more significant at MIR wavelengths, 
where the $N$-band flux in the ID+OD model is dominated by
contributions from the inner and outer disk (resulting in much too
compact emitting structures at 8.7\,$\mu$m and 11.7\,$\mu$m; 
third and fourth panel from top in Fig.~\ref{fig:visprofile-innerouter}), 
while in the ID+GapMaterial+OD model, they are dominated by
optically thin material located in the gap region, resulting in a much
better respresentation of the MIDI and T-ReCS visibilities (Fig.~\ref{fig:visprofile-optthin}).

\section{Discussion}
\label{sec:discussion}

\begin{figure}[tbp]
  \centering
  \includegraphics[angle=0,scale=0.43]{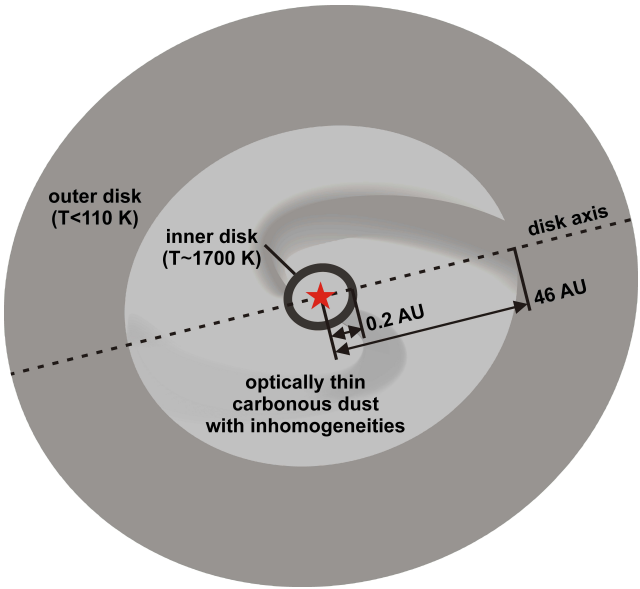}
  \caption{
    Illustration of the V1247\,Ori system, as constrained by our 
    spatially resolved observations (sketch is not to scale).
  }
  \label{fig:sketch}
\end{figure}

Our study provides spatially resolved constraints on the 
structure and physical conditions in a pre-transitional disk (Fig.~\ref{fig:sketch}),
enabling new insights on the disk clearing mechanism in these objects.
We localize a narrow ring of hot material at the expected location of the 
dust sublimation radius, which rules out mechanisms that clear the disk
in an inside-to-outside fashion, such as photoevaporation \citep[e.g.][]{ale07},
magnetorotational instabilities \citep{chi07}, or grain growth \citep{dul05,bir12}.
On the other hand, the observed disk structure supports theories for
dynamical disk clearing by planetary- or sub-stellar-mass companions,
which might be responsible both for the gap-opening and 
for inducing the detected disk asymmetries.
Our NIRC2 and AMBER observations rule out the presence of
stellar-mass companions on scales from $\sim 1$\,mas to a few arcseconds,
with brightness ratios up to $1:100$ (corresponding to companion masses of $0.3\,M_{\sun}$), 
but cannot rule out a very close spectroscopic companion.  
However, such a close stellar-mass companion 
would like also distort the inner disk, either by truncation
or by causing asymmetries, that have not been observed
by our AMBER data.

One of the main results from our study is the detection of significant amounts of 
optically thin material located within the disk gap.
SED-based models of several transitional disks also included an optically thin
component, although in those cases the component has been 
introduced primarily to reproduce spectral features, such as the silicate feature 
(e.g.\ \object{TW\,Hya}, \citealt{cal02}; \object{LkCa\,15}, \citealt{esp11}).
Such spectral features are not present in our V1247\,Ori spectra and 
we require the optically thin component in order to reproduce the
MIR continuum visibilities. 
We find that this component even dominates at MIR wavelengths, which 
is in contrast to \object{HD\,100546} \citep{ben10b,tat11} and 
\object{TW\,Hya} \citep{rat07,ake11}, the other 
transitional/pre-transitional disks that have been extensively studied
with infrared interferometry before.
No optically thin gap material was found in these objects, 
possibly indicating that TW\,Hya and HD\,100546 are already in an later
stage of gap clearing and/or grain growth than V1247\,Ori.
Adopting the dust opacities by \citet{dra84}, we could achieve no satisfactory
fit with silicate dust grains, but require a carbon-dominated dust mineralogy 
for the optically thin material located in the disk gap region.
Of course, this conclusion depends to some degree on the adopted dust properties
and future studies might test also other grain species.

It remains an open question, how this carbon-overabundance has developed,
but it is interesting to note that studies on the $\beta$\,Pictoris planetary system 
have also revealed a carbon-rich composition of the debris material \citep{rob06}.
Also, carbon-rich planets have been predicted by theoretical studies \citep{sea07}
and indications for a carbon-rich interior have already been found for the super-earth 
\object{55\,Cancri\,e} \citep{mad12}.

Using Keck/NIRC2 aperture masking observations, we measured strong non-zero phase signals,
indicating small-scale asymmetries in the brightness distribution.
Similar phase signals have been observed on several other transitional and pre-transitional disks 
(e.g.\ \object{CoKu Tau/4}: \citealt{ire08a};
\object{T\,Cha}; \citealt{hue11};
\object{LkCa\,15}: \citealt{kra12d};
\object{HD142527}: \citealt{bil12}; 
\object{FL\,Cha}: \citealt{cie13}) 
and have often been interpreted as signatures of a close companion.
Fitting the $K'$-band data, we also obtain evidence for a close companion 
($10{\sigma}$ detection significance; Fig.~\ref{fig:phases-NIRC2}, middle column).
However, the direction and amplitude of the asymmetries changes with wavelength,
which suggests that they are not caused by a close companion.
We also rule out that the asymmetries might be caused by the heated wall 
of an inclined centro-symmetric disk, as proposed by \citet{cie13},
as such asymmetries would be consistently directed along the minor axis of the disk.
Instead, our observations reveal more complex, radially extended structures,
whose detailed geometry is not sufficiently resolved by our aperture masking observations.
We speculate that the asymmetries might be caused by spiral-arm features, similar to the
ones discovered with Subaru/HiCIAO coronagraphic observations 
in the pre-transitional disks of \object{HD\,135344B} \citep{mut12} and 
\object{HD\,36112} \citep[=MWC\,758,][]{gra13}
on $\sim 5\times$ larger angular scales.
In both cases, the asymmetries might resemble spiral density waves that are 
induced by orbiting sub-stellar mass companions \citep{gol79},
although alternative scenarios, such as gravitational instabilities \citep[e.g.][]{dur07}
or accretion streams onto forming planets \citep[e.g.][]{zhu12}
have also been proposed.

A major open question concerns the physical cause of the deep UX\,Ori-like 
occultation events that were reported by \citet{cab10}.
The detected two main events were separated by about 14~months and lasted for about two weeks each.
Outsides these isolated events, the photometry of V1247\,Ori was very stable on a 
level below 0.02\,mag, which suggests that the occultations were not related to photospheric activity.
\citet{cab10} proposed that the photometric events might be caused by disk material passing 
the line-of-sight, assuming that the disk is oriented close to edge-on.
At first sight, this scenario seems in contradiction to our inclination measurement of {$i=31 \pm 8$\deg}
(Sect.~\ref{sec:resultsdiskMIR}), suggesting that the disks is seen closer to face-on than edge-on.
However, it should be noted that this estimate has been obtained at MIR wavelengths for the outer
disk regions, while the $uv$-coverage of our NIR data is still insufficient to
obtain an independent estimate for the inclination angle of the inner disk.
Therefore, we are not able to rule out that the inner disk might be considerably inclined or warped with 
respect to the outer disk.  Hypothetical planets on highly inclined orbits in the inner astronomical unit
\citep{sim12} might shepard, deform, and tilt the inner disk with respect 
to the outer disk \citep{muz09}, which might result in the observed occultation events.
Rotation-induced projection affects of this inclined disk might also be responsible for
the detected continuum variability, which peaks at NIR wavelengths and appears
on time scales of several months to years (Sect.~\ref{sec:resultsdiskSED}), 
matching the orbital period of the inner disk.
Alternatively, the occultation events might be related to the 
asymmetric structures that we have detected on scales of $\sim$ 15--40\,AU,
for instance through a large, vertically extended disk warp that intersects
the line-of-sight to the stellar photosphere sporadously.
However, both for time scale arguments and hydrodynamic arguments,
we consider the aforementioned scenario, which relates the occultation
events with the inner disk, more likely.

\section{Conclusions}
\label{sec:conclusions}

We have presented coordinated near- and mid-infrared interferometric observations
of a star with a pre-transitional disk, which has allowed us to investigate the 
disk structure of one of these intriguing objects in unprecedented detail.

Our observations have targeted the previously sparsely studied pre-transitional disk
around V1247\,Ori and resolved a narrow, optically thick inner disk component located at 
0.18\,AU from the star, which dominates the near-infrared emission.
This hot inner disk contributes less than 20\% to the mid-infrared (8--13\,$\mu$m) regime,
while the largest fraction arises from optically thin material located in the gapped disk region 
from $\sim 0.2$ to 46\,AU.
This strong mid-infrared continuum emission from carbonous dust grains located inside the 
gap region has not been predicted by the current class of SED-based disk models.
Multi-wavelength interferometric observations on a larger sample of transitional and 
pre-transitional disks will be necessary to investigate how common this optical thin 
emission component occurs and whether it might be indicative of a 
particularly early stage of disk clearing.
The organic bands (hydrocarbon features) originate from similar stellocentric radii
as this optically thin material.

Our aperture masking observations reveal highly significant asymmetries in the 
$H$-, $K'$-, and $L'$-band.
The direction and amplitude of the asymmetries changes with wavelength,
which leads us to reject a companion origin and to exclude that the asymmetries
are induced by a heated disk wall, as proposed in the scenario by \citet{cie13}.
Instead, we propose that the asymmetries are related to strong density inhomogenities 
in the gap region, possibly caused by the dynamical interaction of the gap material
with the gap-opening body/bodies.
Our results illustrate the highly dynamical and complex nature encountered in the
inner gap regions and demonstrate the need for further high-angular resolution imaging studies.  
We would like to emphasize that the observed disk asymmetries can be easily misinterpreted 
as companion signals and multi-epoch, multi-wavelength studies are essential for ruling out false detections.

With a mass of $1.9\,M_{\sun}$, V1247\,Ori is considerably more massive 
than other pre-transitional disks, approaching the intermediate-mass regime.
Studying planet formation in such intermediate-mass systems (spectral type A or B)
is particular interesting,
as abundance studies suggest that the planet formation efficiency 
increases towards higher stellar masses \citep{joh07b}.
Besides being more abundant than those around solar-type stars, 
giant planets around early-type stars appear to follow also a different 
major-axis distribution, with a significant deficit in the inner 1--2\,AU, 
likely because of the different disk environments in which they formed. 
Multi-wavelength interferometric observations on V1247\,Ori and other 
transitional disks now provide the exciting opportunity to study the
very early evolutionary phases of such systems and to explore the impact
of planet formation on the disk environment directly.
Such observational evidence is essential to test planet formation theories 
and to link our knowledge of planet formation and disk evolution 
to the planetary system demographics observed in main-sequence systems.

\acknowledgments

We thank the referee, Jos\'{e} Caballero, for his detailed report,
which helped to improve the presentation of this paper.
This work was done in part under contract with the California
Institute of Technology (Caltech), funded by NASA through the Sagan Fellowship
Program (S.K. and C.E. are Sagan Fellows).
Data presented herein were obtained at the W.\ M.\ Keck Observatory from telescope time 
allocated to the National Aeronautics and Space Administration through the agency’s 
scientific partnership with the California Institute of Technology and the University of California. 
The Observatory was made possible by the generous financial support of the W.\ M.\ Keck Foundation.
The authors wish to recognize and acknowledge the very significant cultural role and reverence 
that the summit of Mauna Kea has always had within the indigenous Hawaiian community. 
We are most fortunate to have the opportunity to conduct observations from this mountain.
This work was supported in part by the Aerospace Corporation’s Independent Research and 
Development (IR\&D) program.  This work was supported by NASA ADP grant  NNX09AC73G.

{\it Facilities:} \facility{ESO:3.6m}, \facility{Gemini:South}, \facility{IRTF}, \facility{Keck:II}, \facility{Keck:Interferometer}, \facility{VLTI}

\bibliographystyle{apj}
\bibliography{v1247ori}

\end{document}